\documentclass[twocolumn]{aastex631}

\usepackage{float,graphicx,amsmath,multirow}
\usepackage[version=4]{mhchem}
\usepackage{booktabs}
\usepackage{enumitem}
\usepackage{wrapfig}
\usepackage{threeparttable}
\usepackage{comment}
\usepackage{upgreek}

\begin{document}

\title{Discovery of the 7-ring PAH Cyanocoronene (\ce{C24H11CN}) in GOTHAM Observations of TMC-1}

\author[0000-0002-0332-2641]{Gabi Wenzel}
\affiliation{Department of Chemistry, Massachusetts Institute of Technology, Cambridge, MA 02139, USA}
\affiliation{Center for Astrophysics \textbar{} Harvard \& Smithsonian, Cambridge, MA 02138, USA}
\author[0009-0009-5979-9041]{Siyuan Gong}
\affiliation{Department of Chemistry, Massachusetts Institute of Technology, Cambridge, MA 02139, USA}
\author[0000-0003-2760-2119]{Ci Xue}
\affiliation{Department of Chemistry, Massachusetts Institute of Technology, Cambridge, MA 02139, USA}
\author[0000-0003-0304-9814]{P. Bryan Changala}
\affiliation{JILA, University of Colorado Boulder and National Institute of Standards and Technology, Boulder, CO 80309, USA}
\affiliation{Department of Physics, University of Colorado Boulder, Boulder CO 80309, USA}
\author[0009-0008-1171-278X]{Martin S. Holdren}
\affiliation{Department of Chemistry, Massachusetts Institute of Technology, Cambridge, MA 02139, USA}
\author[0000-0001-8134-5681]{Thomas H. Speak}
\affiliation{Department of Chemistry, University of British Columbia, Vancouver, BC, Canada}
\author[0009-0005-1773-8460]{D. Archie Stewart}
\affiliation{Department of Chemistry, Massachusetts Institute of Technology, Cambridge, MA 02139, USA}
\author[0000-0001-5020-5774]{Zachary T. P. Fried}
\affiliation{Department of Chemistry, Massachusetts Institute of Technology, Cambridge, MA 02139, USA}
\author[0009-0002-6372-9926]{Reace H. J. Willis}
\affiliation{Department of Chemistry, University of British Columbia, Vancouver, BC, Canada}


\author[0000-0003-4179-6394]{Edwin A. Bergin}
\affiliation{Department of Astronomy, University of Michigan, Ann Arbor, MI 48109, USA}
\author[0000-0003-0799-0927]{Andrew M. Burkhardt}
\affiliation{Department of Earth, Environment, and Physics, Worcester State University, Worcester, MA 01602, USA}
\author[0000-0002-4593-518X]{Alex N. Byrne}
\affiliation{Department of Chemistry, Massachusetts Institute of Technology, Cambridge, MA 02139, USA}
\author[0000-0001-6752-5109]{Steven B. Charnley} 
\affiliation{Astrochemistry Laboratory and the Goddard Center of Astrobiology, Solar System Exploration Division, NASA Goddard Space Flight Center, Greenbelt, MD 20771, USA }
\author[0000-0002-6667-7773]{Andrew Lipnicky}
\affiliation{National Radio Astronomy Observatory, Charlottesville, VA 22903, USA}
\author[0000-0002-8932-1219]{Ryan A. Loomis}
\affiliation{National Radio Astronomy Observatory, Charlottesville, VA 22903, USA}
\author[0000-0002-5171-7568]{Christopher N. Shingledecker}
\affiliation{Department of Chemistry, Virginia Military Institute, Lexington, VA 24450, USA}


\author[0000-0002-0850-7426]{Ilsa R. Cooke}
\affiliation{Department of Chemistry, University of British Columbia, Vancouver, BC, Canada}
\author[0000-0001-9479-9287]{Anthony J. Remijan}
\affiliation{National Radio Astronomy Observatory, Charlottesville, VA 22903, USA}
\author[0000-0001-9142-0008]{Michael C. McCarthy}
\affiliation{Center for Astrophysics \textbar{} Harvard \& Smithsonian, Cambridge, MA 02138, USA}
\author[0000-0003-2970-9817]{Alison E. Wendlandt}
\affiliation{Department of Chemistry, Massachusetts Institute of Technology, Cambridge, MA 02139, USA}
\author[0000-0003-1254-4817]{Brett A. McGuire}
\affiliation{Department of Chemistry, Massachusetts Institute of Technology, Cambridge, MA 02139, USA}
\affiliation{National Radio Astronomy Observatory, Charlottesville, VA 22903, USA}




\correspondingauthor{Gabi Wenzel, Brett A. McGuire}
\email{gwenzel@mit.edu, brettmc@mit.edu}

\begin{abstract}

We present the synthesis and laboratory rotational spectroscopy of the 7-ring polycyclic aromatic hydrocarbon (PAH) cyanocoronene (\ce{C24H11CN}) using a laser-ablation assisted cavity-enhanced Fourier transform microwave spectrometer. A total of 71 transitions were measured and assigned between 6.8--10.6\,GHz. Using these assignments, we searched for emission from cyanocoronene in the GBT Observations of TMC-1: Hunting Aromatic Molecules (GOTHAM) project observations of the cold dark molecular cloud TMC-1 using the 100\,m Green Bank Telescope (GBT). We detect a number of individually resolved transitions in ultrasensitive X-band observations and perform a Markov Chain Monte Carlo analysis to derive best-fit parameters, including a total column density of $N(\ce{C24H11CN}) = 2.69^{+0.26}_{-0.23} \times 10^{12}\,\mathrm{cm}^{-2}$ at a temperature of $6.05^{+0.38}_{-0.37}\,$K. A spectral stacking and matched filtering analysis provides a robust 17.3$\,\sigma$ significance to the overall detection. The derived column density is comparable to that of cyano-substituted naphthalene, acenaphthylene, and pyrene, defying the trend of decreasing abundance with increasing molecular size and complexity found for carbon chains. We discuss the implications of the detection for our understanding of interstellar PAH chemistry and highlight major open questions and next steps.
\end{abstract}



\section{Introduction}
\label{sec:intro}

Polycyclic aromatic hydrocarbons (PAHs) are a class of molecules thought to sequester a substantial portion (10--25\,\%) of the interstellar carbon budget~\citep{Tielens:2008:289,Chabot:2020:17,Dwek:1997:565,Habart:2004:179}, contribute to interstellar \ce{H2} formation both as catalytic surfaces~\citep{Mennella:2012:L2,Thrower:2012:3,Rauls:2008:531,LePage:2009:274} and through H abstraction~\citep{Bauschlicher:1998:L125,Boschman:2015:A72}, serve as charge balance carriers~\citep{Carelli:2012:3643,Bakes:1998:258}, and contribute to neutral gas heating in interstellar sources~\citep{Berne:2022:A159}, among numerous other roles. Since the mid-1980s, the unidentified infrared bands (UIRs), broad emission features in the mid-infrared range observed toward many astronomical objects which are classified as photon-dominated regions (PDRs) of the interstellar medium (ISM), have been widely attributed to vibrational modes of electronically excited PAHs~\citep{Leger:1984:279, Allamandola:1985:L25} --- for this reason, these bands are also commonly referred to as the aromatic infrared bands (AIBs) --- although no individual PAH carrier has been identified via its infrared emission~\citep{Tielens:2008:289}. Definitive evidence for the presence of PAHs outside the solar system, however, was obtained in 2021 with the discoveries of 1- and 2-cyanonaphthalene (\ce{C10H7CN}) via radio astronomy~\citep{McGuire:2021:1265}. The detections were made in the cold, starless molecular cloud TMC-1 with the 100\,m Robert C. Byrd Green Bank Telescope (GBT) as part of the GBT Observations of TMC-1: Hunting Aromatic Molecules (GOTHAM) large program.  Shortly thereafter, a third PAH, indene (\ce{C9H8}; \citealt{Burkhardt:2021:L18,Cernicharo:2021:L15}) was discovered both by GOTHAM and the Yebes 40\,m telescope Q-band Ultrasensitive Inspection Journey to the Obscure TMC-1 Environment (QUIJOTE) project, followed the next year with the detection of 2-cyanoindene (\ce{C9H7CN}; \citealt{Sita:2022:L12}) by GOTHAM. The detections of 1- and 5-cyanoacenaphthylene (\ce{C12H7CN}) were then reported by \citet{Cernicharo:2024:L13}. 

Most recently, 1-, 2-, and 4-cyanopyrene (\ce{C16H9CN}) were discovered in the GOTHAM observations~\citep{Wenzel:2024:810,Wenzel:2025:262}. These four-ring, peri-condensed PAHs were the largest species detected with radio astronomy to date. From these observations, it was estimated that the unsubstituted (bare) pyrene (\ce{C16H10}) itself may account for as much as 0.1\,\% of the carbon budget~\citep{Wenzel:2024:810}. Notably, the derived column densities of the known PAHs in TMC-1 are all similar: the cyanonaphthalenes, cyanoacenaphthylenes, and cyanopyrenes all lie within a factor of $\sim$2 of each other between $0.75-1.5\times10^{12}\,\mathrm{cm}^{-2}$. (2-cyanoindene is somewhat lower at $0.2\times10^{12}\,\mathrm{cm}^{-2}$.) This is in stark contrast to the generally observed column density trends with carbon atom numbers within a molecular family~\citep{Loomis:2021:188,Siebert:2022:21}. Given these results, we considered whether yet larger PAHs may be detectable with similar column density in TMC-1.

\begin{figure}
    \centering
    \includegraphics[width=0.75\columnwidth]{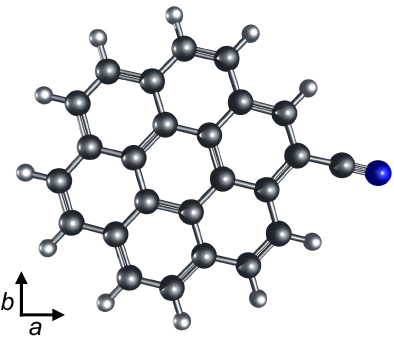}
    \caption{Optimized geometry of cyanocoronene, \ce{C24H11CN}, in its principal axis system spanned by the vectors $a$ and $b$. The highly symmetric coronene has 12 identical carbon sites for \ce{CN}-substitution that all result in the same cyanocoronene. Its permanent electric dipole moment was calculated to be $\mu_a = 5.67\,\mathrm{D}$ and $\mu_b = 0.59\,\mathrm{D}$.}
    \label{fig:cn-coro-geom}
\end{figure}

Coronene (\ce{C24H12}) is often described in the literature as the ``prototypical" peri-condensed (compact) PAH, 
renowned for its structural, chemical, and spectral properties, making it a model for understanding larger PAHs in terrestrial and extraterrestrial environments. Its $D_{6h}$ symmetry and highly conjugated $\pi$-electron system enhance its aromatic character and stability, making it energetically favorable and distinguishing it from simpler PAHs like naphthalene and anthracene~\citep{Clar:1964:,Clar:1983:49}. Indeed, the compellingly close agreement with small shifts (${\leq}0.6\,\mathrm{\upmu m}$) between the computed vibrational features of coronene and the observed AIBs was used by~\citet{Leger:1984:279} in their original suggestion that PAHs were the primary carriers. Due to its role as a ``model PAH,'' many astrochemical studies have focused on coronene and its derivatives, clusters, fragments, and charge states. Coronene is efficiently produced at high temperatures due to its thermodynamic stability ~\citep{Stein:1978:566,Hudgins:1995:3033}, is the largest PAH found in the Murchison meteorite~\citep{Sephton:2004:1385,Sabbah:2017:34}, and among the PAHs identified in return samples from asteroid Ryugu~\citep{Sabbah:2024:e20240010}. Its infrared spectrum is known for its neutral form at varying temperatures~\citep{Joblin:1995:835} and for its cationic form in ion trap experiments~\citep{Oomens:2001:L99}. The infrared spectrum of protonated coronene (\ce{C24H13+}) reveals it to be a key species among the potential carriers of UIR bands (or AIBs)~\citep{Dopfer:2011:103, Bahou:2014:1039,Cazaux:2016:19835}. The destruction of coronene may produce key reactive intermediates involved in the formation of smaller complex organic molecules (COMs) and fullerenes~\citep{Gatchell:2021:6646,Chen:2020:A103, Panchagnula:2024:18557}. However, no acetylene loss, one of the major fragmentation channels of smaller PAHs, has been observed for coronene cations in experiments resembling PDRs with vacuum ultraviolet (VUV) photons up to $20\,\mathrm{eV}$~\citep{Zhen:2016:113,Joblin:2020:062002}, further emphasizing its extreme stability. Coronene's role in the context of the diffuse interstellar bands (DIBs) has been debated. Dehydrogenated, protonated, and cationic coronene were proposed as potential carriers of DIBs~\citep{Pathak:2008:L10,Malloci:2008:1183}, but these hypotheses were disproven~\citep{Garkusha:2011:10972,Useli-Bacchitta:2010:16,Hardy:2017:37}. Nevertheless, some distribution of PAHs remain promising DIB carrier candidates due to their close relation to the fullerene family~\citep{Campbell:2015:322}. 

As with many PAHs, coronene possesses no permanent electric dipole moment and thus no pure-rotational spectrum. A search for its \ce{CN}-substituted variant is therefore reasonable, and indeed, experimental evidence has previously shown that the addition of \ce{CN} to coronene may proceed readily under interstellar conditions~\citep{Bernstein:2002:1115}, although we note that the study in question was specifically in the condensed phase. Here, we present a combined computational and laboratory study of the rotational spectrum of cyanocoronene (\ce{C24H11CN}) and its subsequent discovery in TMC-1.

\section{Synthesis}
\label{sec:synthesis}

Cyanocoronene, the \ce{CN}-derivative of the highly symmetric PAH coronene (see Fig.~\ref{fig:cn-coro-geom}), is not commercially available. To measure the laboratory rotational spectrum of cyanocoronene, we synthesized it via the route described in detail in Appendix~\ref{app:synthesis}, which was partially based on previous work~\citep{Dale:2006:4500,Hyodo:2017:3005}. Briefly, coronene (\ce{C24H12}) was purchased from Ambeed Inc. (purity ${\sim}98\,\%$) and used as the starting material to prepare formylcoronene (\ce{C24H11CHO}). Formylcoronene further reacted via formylcoronene-oxime (\ce{C24H11CHNOH}) to cyanocoronene (\ce{C24H11CN}), a yellow solid, with an approximate yield of 59\,\%. 

\section{Laboratory Spectroscopy}
\label{sec:spectroscopy}

The geometry of cyanocoronene (see Fig.~\ref{fig:cn-coro-geom}) was first optimized at the $\omega$B97X-D4/def2-TZVPP level~\citep{Chai:2008:084106,Caldeweyher:2017:034112,Weigend:2006:1057} in ORCA 5.0.4~\citep{Neese:2022:e1606} and re-optimized at the B3LYP/aug-cc-pVTZ level \citep{Becke:1993:5648,Dunning:1989:1007a,Kendall:1992:6796,Davidson:1996:514} in Gaussian 16~\citep{g16} , which has recently been shown to perform well when obtaining rotational constants of PAHs~\citep{Neeman::e202401012}. We further benchmarked its performance using the three cyanopyrene isomers previously studied (see Appendix~\ref{app:benchmark} and Table~\ref{tab:benchmark_b3lyp}). Together with quartic centrifugal distortion constants derived from the harmonic force field (see Table~\ref{tab:constants}), these theoretically calculated spectroscopic constants formed the basis of our laboratory search.

\begin{table}[ht!]
    \centering
    \footnotesize
    \caption{Rotational constants of cyanocoronene in the A-reduced III$^l$ representation. $N^\mathrm{fit}_\mathrm{lines}$ and $N^\mathrm{unique}_\mathrm{lines}$ refer to number of distinct transitions in the fit and number of unique transition frequencies in the fit, respectively. See Table~\ref{tab:transitions} for the line list.}    
    \begin{tabular}{l r r}
    \toprule
    Parameter               &  B3LYP     &   Experimental\tablenotemark{\footnotesize{a}}\\
                   &  aug-cc-pVTZ     &    \\
    \midrule

    $A$ (MHz)               &  335.393               &  333.852989(249)
    \\
    $B$ (MHz)               &  222.035             &   221.2700880(838)  \\
    $C$ (MHz)               &  133.594           &   133.1081015(122) \\
    $\Delta_J$ (Hz)         &  0.669                  &   [0.669]  \\
    $\Delta_{JK}$ (Hz)      &  -0.617                 &  [-0.617]    \\
    $\Delta_K$ (Hz)         &  0.011                 &  [0.011]  \\
    $\delta_J$ (Hz)        &  -0.120   & [-0.120] \\ 
    $\delta_K$ (Hz)        &  -0.494  & [-0.494] \\
    & \\
    $N^\mathrm{fit}_\mathrm{lines}$ & & 71 \\
    $N^\mathrm{unique}_\mathrm{lines}$ & & 38 \\
    $\sigma_\mathrm{fit}$ (kHz) & & 2.595 \\
    $(J, K_{a})_\mathrm{max}$ & & {$(39, 10)$} \\
    \bottomrule 
    \end{tabular}
    \tablenotetext{a}{Values in parentheses are 1$\,\sigma$ uncertainties in units of the last digit. Values in brackets were not determinable and fixed to the theoretically calculated constants.}
    \label{tab:constants}
\end{table}

Details of the sample preparation and spectroscopic measurements are given in \citet{Wenzel:2024:810}. Briefly, a hydraulic press was used to compress approximately 600\,mg of solid sample into a cylindrical sample rod, which was then mounted in a rotating stage located downstream of a pulsed valve backed with neon carrier gas. The rod was then ablated using the second harmonic at 532\,nm of a Nd:YAG laser to inject the sample into the gas phase, where it was entrained in the neon and supersonically expanded into the cavity of a Balle-Flygare type Fourier transform microwave (FTMW) spectrometer~\citep{Balle:1981:33,Crabtree:2016:124201}. At the $\sim$2\,K rotational temperatures generated by the supersonic expansion, the strongest transitions of cyanocoronene fall in the very lowest end of the operational range of the instrument, between 6--8\,GHz.

In total, we were able to measure and assign 71 transitions (38 unique lines) of cyanocoronene (see Table~\ref{tab:transitions}). We used \textsc{SPCAT/SPFIT}~\citep{Pickett:1991:371} to determine the spectroscopic constants of cyanocoronene by least-squares fitting in a Watson A reduction (III$^l$ representation). The asymmetry parameter, $\kappa = -0.12$, places cyanocoronene far from either the prolate or oblate symmetric-top limits. Centrifugal distortion constants were fixed to the theoretically calculated values. The $^{14}$N nuclear electric quadrupole coupling was neglected entirely because no splittings could be resolved for the high-$J$ transitions measured (see Table~\ref{tab:transitions}). The best-fit rotational constants, $A$, $B$, and $C$, are listed in Table~\ref{tab:constants} and with a mean absolute percentage error (MAPE) of 0.39\,\% in excellent agreement with the calculated values.

\section{Observations}
\label{sec:obs}

\begin{figure}[htb!]
    \centering
    \includegraphics[width=\columnwidth]{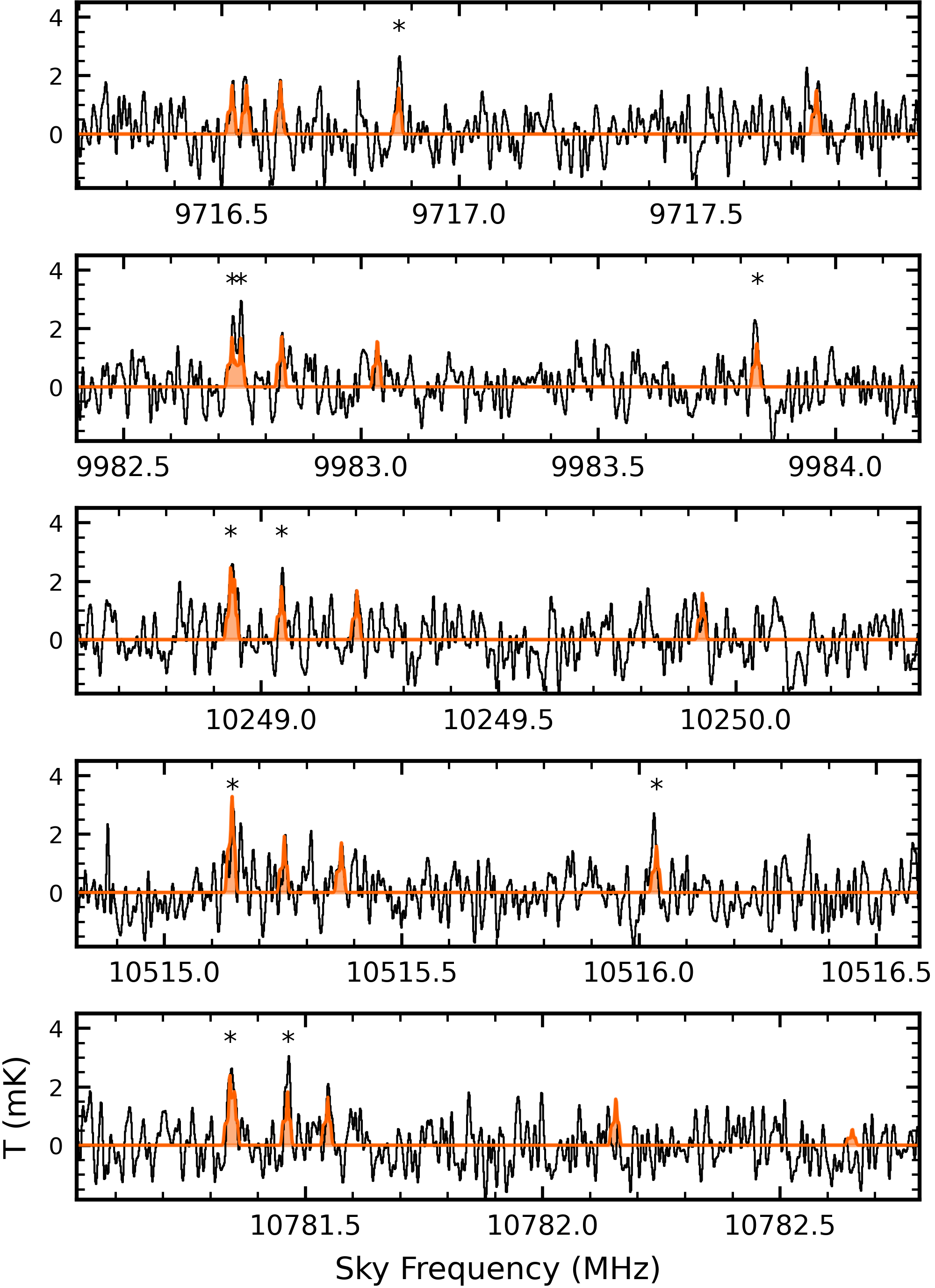}
    \caption{GOTHAM spectra smoothed with a 10-channel Hanning window to a resolution of 14\,kHz (black) overlaid with the spectra of cyanocoronene (orange) simulated using the MCMC-derived parameters given in Table~\ref{tab:mcmc}. Lines with SNR$\,>3\,\sigma$ are marked with asterisks.}
    \label{fig:indiv_lines}
\end{figure}

The observations of TMC-1 were acquired on the 100\,m Robert C. Byrd Green Bank Telescope (GBT) as part of the GOTHAM project~\citep{McGuire:2020:L10}. Details of the observing strategy and data reduction procedures are provided elsewhere~\citep{McGuire:2020:L10,McGuire:2021:1265,Sita:2022:L12}. For this analysis, we used observations from the fifth data reduction which includes advances in artifact and radio-frequency interference removal, atmospheric opacity corrections, and baseline removal~\citep{pipeline}. Briefly, we target the ``cyanopolyyne peak" of TMC-1 centered at $\alpha_\mathrm{J2000}$ = 04\fh41\fm42.5\fs, $\delta_\mathrm{J2000}$ = +25\arcdeg41\arcmin26.8\arcsec. The GOTHAM observations cover 29\,GHz of bandwidth between 4--36\,GHz, limited by gaps in receiver coverage, at a resolution of 1.4\,kHz with a root-mean-squared (RMS) noise level between 3.8--15.0\,mK across most of the observed frequency range. Flux calibration was achieved with switched noise-diode measurements, resulting in an estimated antenna temperature accuracy of $10-20\,\%$. In the case of cyanocoronene, the strongest transitions contributing to the overall detection fall in the C-, X-, and Ku-bands. As discussed below, the observation of individual transitions of cyanocoronene was enabled by our ultra-sensitive observations in X-band between 9.38--10.96\,GHz (GBT Project 24A-124). Combining this deep X-band and previous GOTHAM observations, we reach an RMS of 1.5\,mK in this frequency range. While we conduct the analysis described in \S\ref{sec:astroanalysis} on the full-resolution spectra, we also smoothed the data with a 10-channel Hanning window to a resolution of 14\,kHz to increase the signal-to-noise ratio (SNR) on individual transitions and better visualize the detection (see Fig.~\ref{fig:indiv_lines}).

\section{Astronomical Analysis}
\label{sec:astroanalysis}

\begin{figure*}[htb!]
    \centering
    \includegraphics[width=0.49\textwidth]{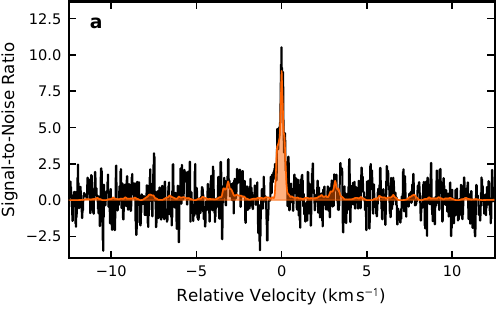}
    \includegraphics[width=0.49\textwidth]{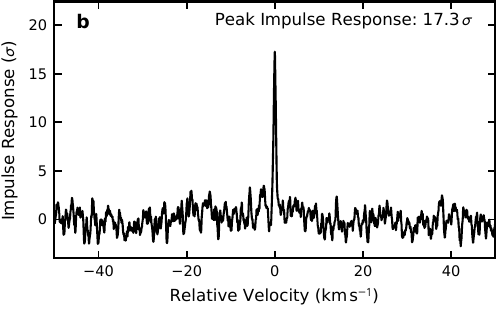}
    \caption{(a) Velocity-stacked spectra and (b) matched filter analysis results of cyanocoronene spectra in the GOTHAM data generated using the methodologies outlined in \citet{Loomis:2021:188}.
    }
    \label{fig:stack_filter}
\end{figure*}

Due to the low RMS noise in our X-band data, the brightest SNR lines of cyanocoronene at TMC-1 conditions ($v_\mathrm{LSR} \sim 5.8\,\mathrm{km\,s^{-1}}$, $T_\mathrm{ex} \sim 5-7$\,K) fall in the frequency range of $8-12$\,GHz. A search for cyanocoronene in the GOTHAM data yielded several individually resolved rotational transitions close to the RMS noise, particularly near 10.515\,GHz (see Fig.~\ref{fig:indiv_lines}). We performed a Markov Chain Monte Carlo (MCMC) analysis adopted from previous work~\citep{McGuire:2021:1265, Loomis:2021:188} to derive physical parameters that best describe the emission of cyanocoronene. This approach provides an inference by conditioning the data on priors (see Table~\ref{tab:MCMCpriors}) and sampling posterior distributions. The uncertainty of observations is computed as the quadrature sum of the local RMS noise and an additional $20\,\%$ systematic uncertainty. The parameters derived from the MCMC analysis are listed in Table~\ref{tab:mcmc}, where the velocities in the local standard of rest, $v_\mathrm{LSR}$, and excitation temperature, $T_\mathrm{ex}$, are consistent with prior detections in TMC-1~\citep{McCarthy:2021:176,Sita:2022:L12}. The derived uncertainties in each parameter reflect the posterior probability distribution.

From our MCMC analysis, the total column density was derived as the sum of the column densities of all four Doppler components, yielding a value of $N(\ce{C24H11CN}) = 2.69^{+0.26}_{-0.23} \times 10^{12}\,\mathrm{cm^{-2}}$ and an $T_\mathrm{ex} = 6.05^{+0.38}_{-0.37}\,$K. This value of $T_\mathrm{ex}$ is consistent with other PAHs detected in TMC-1 \citep{McGuire:2021:1265, Cernicharo:2024:L13, Wenzel:2024:810, Wenzel:2025:262}. To quantify the significance of our cyanocoronene detection, we performed a velocity-stack and matched filtering analysis (as described in~\citealt{Loomis:2021:188,McGuire:2021:1265,Wenzel:2024:810}) of the 100 brightest SNR lines of cyanocoronene, resulting in a positive detection with a high confidence level of 17.3$\,\sigma$ in the matched filter response (see Fig.~\ref{fig:stack_filter}).

\section{Discussion}
\label{sec:discussion}

\begin{figure}[hb!]
    \centering
    \includegraphics[width=0.875\columnwidth]{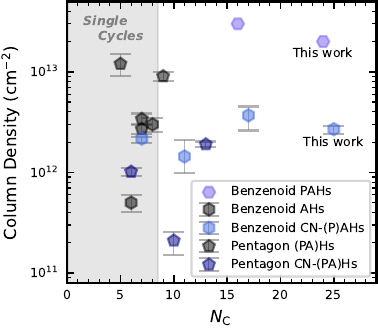}
    \caption{Comparison of the derived column densities of the cyclic hydrocarbon (Hs) population in TMC-1 vs. number of carbon atoms, $N_\mathrm{C}$. We distinguish between single and multiple cycles, pentagon-containing (PA)Hs, solely hexagon-containing (P)AHs, and their \ce{CN}-derivatives. Purple hexagons refer to pure benzenoid PAHs and their column densities were estimated from their \ce{CN}-derivatives (see text and Appendix~\ref{app:mesmer} for details). Black hexagons are pure single benzenoid cycles (aromatic hydrocarbons, AHs). Light blue hexagons correspond to \ce{CN}-substituted benzenoid (P)AHs; black and dark blue pentagons are pure and \ce{CN}-substituted pentagonal (PA)Hs, respectively. Column density values and the references they were taken from are reported in Table~\ref{tab:CDcomp}.}
    \label{fig:CDcomp}
\end{figure}

\begin{figure}[ht!]
    \centering
    \includegraphics[width=\columnwidth]{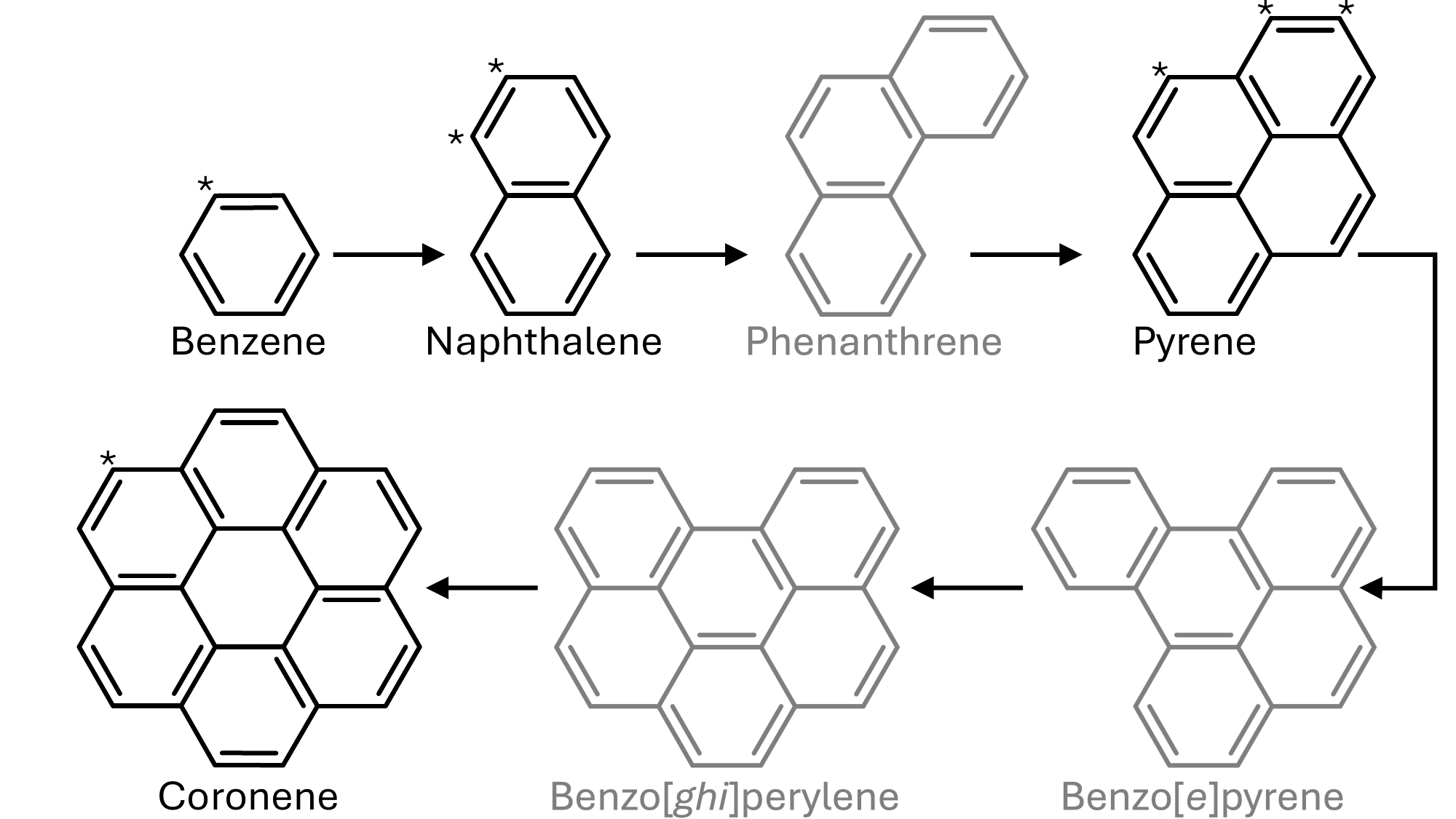}
    \caption{First seven members of the most thermodynamically favorable, high-temperature (P)AH polymerization route, adapted from \citet{Stein:1978:566} and \citet{Hudgins:1995:3033}. Note that this is not a reaction scheme. Those whose \ce{CN}-derivatives have been detected in TMC-1 are depicted in black with their substitution sites marked by asterisks. Those in grey have not (yet) been detected or searched for in TMC-1. See text for details.}
    \label{fig:thermoPAHs}
\end{figure}

Based on the assumption that \ce{CN} addition to aromatic double bonds is barrierless~\citep{balucani_formation_2000,woon_quantum_2009,Cooke:2020:L41}, we use the detected \ce{CN}-derivatives as proxies to infer the presence of their unsubstituted parent PAHs in TMC-1. Analogous to our estimate of the column density of pyrene in TMC-1~\citep{Wenzel:2025:262}, we use the \ce{CN}/\ce{H} ratio to derive an approximate column density of coronene of $N(\ce{C24H12}) \approx 2 \times 10^{13}\,\mathrm{cm}^{-2}$ (see Appendix~\ref{app:mesmer} for details). Cyanocoronene, with its 24 carbon atoms (excluding the \ce{CN} group), is by far the largest PAH found to date by radio telescopes in the ISM. Together with the previous detections of cyclic (PA)Hs in TMC-1, which have an approximately flat distribution in column density with increasing size (see Fig.~\ref{fig:CDcomp}), this result challenges our understanding of the chemistry at play in the dense ISM. Although no low-temperature bottom-up gas-phase formation routes are known for compact medium-sized PAHs such as pyrene, it is remarkable that the detected \ce{CN}-derivatives of the mono- and polycyclic aromatic hydrocarbons benzene, naphthalene, pyrene, and now coronene, are all members of the most thermodyanmically stable PAHs formed by the high-temperature polymerization route (see Fig.~\ref{fig:thermoPAHs}; \citealt{Stein:1978:566,Hudgins:1995:3033}). This polymerization route builds up large, highly peri-condensed PAHs with low \ce{H}/\ce{C} ratios by adding one ring at a time, e.g., via the \ce{H}-abstraction acetylene (\ce{C2H2}) addition (HACA) mechanism~\citep{Reizer:2022:132793}.

Considering environments of circumstellar envelopes, \citet{Zhao:2018:413} proposed a high-temperature gas-phase formation route for pyrene, starting from 4-phenanthrenyl radical. Indeed, phenanthrene is also a member of the most thermodynamically stable PAH polymerization route (see Fig~\ref{fig:thermoPAHs}). However, cursory searches for its \ce{CN}-derivative 9-cyanophenanthrene, whose rotational spectrum is known~\citep{McNaughton:2018:5268}, in our GOTHAM observations have not yet been successful. This could be due to the fact that the spectroscopy of only one of the five possible cyanophenanthrene isomers is known, while others might be more abundant. However, it is consistent with the findings by \citet{Zeichner:2023:1411} that the 3-ring (linear or non-compact) PAH species identified in return samples from asteroid Ryugu were formed at high temperatures, while 2- and 4-ring PAH species were formed via a kinetically controlled route at low temperatures (${\sim}10\,\mathrm{K}$), in line with our findings in TMC-1. To our knowledge, neither the other two members of the series, benzo[\textit{e}]pyrene and benzo[\textit{ghi}]perylene, nor their \ce{CN}-substituted derivatives (six possible individual addition sites per species) have yet been characterized by laboratory rotational spectroscopy. Measuring their laboratory spectra will be crucial for searching for these species in TMC-1 and related sources, especially because high-temperature gas-phase formation routes from benzo[\textit{e}]pyrene and benzo[\textit{ghi}]perylene to coronene in circumstellar envelopes have recently been revealed~\citep{Goettl:2023:15443}. Detections of, or upper limit constraints on, such intermediates on the path to larger PAH formation in TMC-1 will be critical for elucidating this new and unexplored cold complex chemistry.

In addition to considering formation routes, however, it is becoming increasingly clear that resilience to the major destruction routes of interstellar molecules likely represents another factor which may help to explain the surprisingly flat abundance of PAHs even out to the size of cyanocoronene. For gas-phase species in molecular clouds, generally, two of the main destruction routes are reactions with ions and depletion onto grains -- both of which have been the subject of very recent studies involving PAHs. For ion-neutral reactions, recent ion-beam storage ring experiments by \citet{stockett_dissociation_2025} and \citet{bull_radiative_2025} indicate that the underlying PAH backbone can be maintained via efficient radiative cooling, thereby opening up the possibility of a kind of chemical ``recycling'' of PAHs that would certainly contribute to the high observed abundance of, e.g., cyanocoronene. Turning to depletion onto grains, \citet{dartois_desorption_2025} conducted experiments on the sputtering yield of solid-phase perylene and coronene bombarded by energetic ions, analogous to the cosmic ray exposure of species in dust-grain ice mantles. They found that such cosmic ray-induced sputtering is efficient under ISM conditions, and even predict a gas-phase fractional abundance of coronene over $10^{-10}$, in agreement with the findings presented here. The incorporation of such findings into astrochemical models therefore represents a very promising means of substantially improving their agreement with astronomical observations.

\section{Conclusions}
\label{sec:conclusions}

We report the interstellar identification of cyanocoronene, a nitrile derivative of the 7-ring PAH coronene, in GOTHAM observations of the cold molecular cloud TMC-1. We derive high column densities of cyanocoronene and its unsubstituted parent, coronene, of $N(\ce{C24H11CN}) = 2.69^{+0.26}_{-0.23}\times 10^{12}\,\mathrm{cm}^{-2}$ and $N(\ce{C24H12}) \approx 2 \times 10^{13}\,\mathrm{cm}^{-2}$, respectively. Cyanocoronene is the largest individual PAH discovered in space and is present in similar column density to the 4-ring PAH cyanopyrene, suggesting an unexplored reservoir of larger PAHs in the ISM. This discovery delivers additional support for the PAH hypothesis and further evidence of the ubiquitous presence of PAHs in space. Comparisons to organics in the Murchison meteorite and asteroid Ryugu suggest a substantial inheritance of PAHs, possibly produced in the cold ($T \sim 10\,$K) conditions that occur ${\sim}$1\,Myr before star birth. They represent a promising source of carbon for forming terrestrial worlds in stellar systems, to which carbon is supplied in the form of solid-state organics~\citep{Li:2021:eabd3632} from their own natal clouds.

\section{Data access \& code}
The raw data of the GOTHAM observations are publicly available in the GBT Legacy Data Archive\footnote{\footnotesize \url{https://greenbankobservatory.org/portal/gbt/gbt-legacy-archive/gotham-data/}}. The code used to perform the analysis is part of the \texttt{molsim} open-source package; an archival version of the code can be accessed at \citet{molsim}. Calibrated and reduced observational data windowed around the reported transitions, the full catalog of cyanocoronene (including quantum numbers of each transition), and the partition functions used in the MCMC analysis are available via Zenodo at \url{https://doi.org/10.5281/zenodo.15150735}.

\facilities{GBT}



\section*{Acknowledgments}

G.W. would like to dedicate this paper to her PhD supervisor, Christine Joblin, whose passion for coronene and astro-PAHs in general has greatly influenced the direction of her research and whose guidance continues to inspire her. The authors thank H. Gupta for assistance in conducting observations. We gratefully acknowledge support from NSF grants AST-1908576, AST-2205126, and AST-2307137. G.W. and B.A.M. acknowledge the support of the Arnold and Mabel Beckman Foundation Beckman Young Investigator Award. Z.T.P.F. and B.A.M. gratefully acknowledge the support of Schmidt Family Futures. I.R.C. acknowledges support from the University of British Columbia and the Natural Sciences and Engineering Research Council of Canada (NSERC). I.R.C. and T.H.S. acknowledge the support of the Canadian Space Agency (CSA) through grant 24AO3UBC14. P.B.C. is supported by NIST. The National Radio Astronomy Observatory is a facility of the National Science Foundation operated under cooperative agreement by Associated Universities, Inc. The Green Bank Observatory is a facility of the National Science Foundation operated under cooperative agreement by Associated Universities, Inc.  

\bibliographystyle{aasjournal}
\bibliography{CyanocoroneneinTMC-1}

\appendix
\restartappendixnumbering

\section{Full Synthesis Route to Cyanocoronene}
\label{app:synthesis}

\begin{figure}[htb!]
    \centering
    \includegraphics[width=0.8\linewidth]{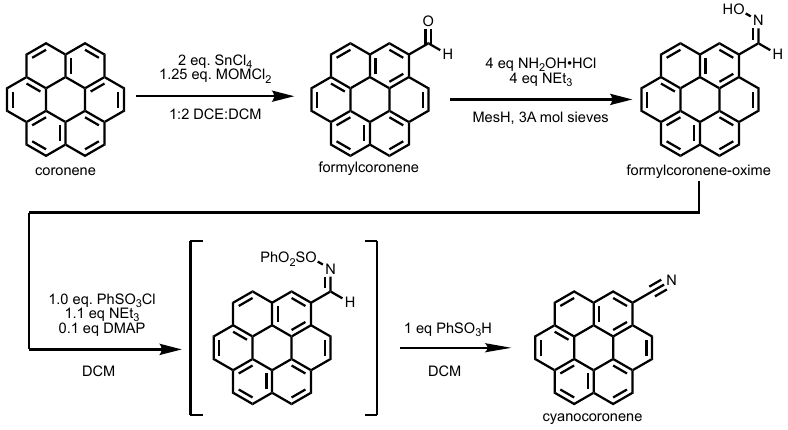}
    \caption{Full synthesis route from coronene (\ce{C24H12}) to cyanocoronene (\ce{C24H11CN}).}
    \label{fig:synthesis}
\end{figure}

The full synthesis route from coronene to cyanocoronene is depicted in Fig.~\ref{fig:synthesis} and described in the following.

\subsection{Synthesis of Formylcoronene}

Formylcoronene was prepared in accordance with the literature~\citep{Dale:2006:4500}. To an oven-dried 500\,mL Schlenk flask was added coronene (Ambeed Inc., 5.00\,g, 16.7\,mmol, 1\,eq). After purging with \ce{N2}, 80\,mL anhydrous 1,2-dichloroethane (DCE) and 160 mL anhydrous dichloromethane (DCM) were added. The flask was then chilled to 0\,$^\circ$C in an ice bath, and anhydrous \ce{SnCl4} (3.9\,mL, 33\,mmol, 2\,eq) was injected into the mixture. Over the course of an hour, dichloromethyl methyl ether (\ce{MOMCl2}, 1.9\,mL, 21\,mmol, 1.25\,eq) was dropwise added into the mixture while the temperature was maintained at 0\,$^\circ$C. The reaction was stirred for an additional 2 hours at 0\,$^\circ$C before warming to 55\,$^\circ$C and held at that temperature for 30 minutes. It was subsequently gradually cooled to room temperature and stirred for 19 hours. The reaction mixture was again cooled to 0\,$^\circ$C, and 100\,mL ice-cold water was added. After stirring for 3 hours at room temperature, the reaction was extracted 5 times with 100\,mL DCM each. Anhydrous \ce{Na2SO4} was added to the organic phase, and the whole crude mixture was subjected to column chromatography using pure DCM to obtain formylcoronene (2.9\,g, 8.8\,mmol, 53\,\% yield) as yellow needle-like crystals. $^{1}$H nuclear magnetic resonance (NMR) (400\,MHz, \ce{CDCl3}): $\delta$ 10.88 (s, 1H), 10.02 (d, $J$ = 8.8\,Hz, 1H), 9.01 (s, 1H), 8.91 - 8.64 (m, 9H). $^{13}$C NMR (126\,MHz, \ce{CDCl3}) $\delta$ 194.39, 136.78, 130.51, 129.16, 128.81, 128.59, 128.09, 127.95, 126.81, 126.79, 126.70, 126.60, 126.55, 126.34, 126.23, 125.68, 124.85, 122.90, 122.78, 122.09, 122.01, 121.72, 121.70 (Peaks may overlap).\\

\begin{figure}
    \centering
    \includegraphics[width=0.8\linewidth]{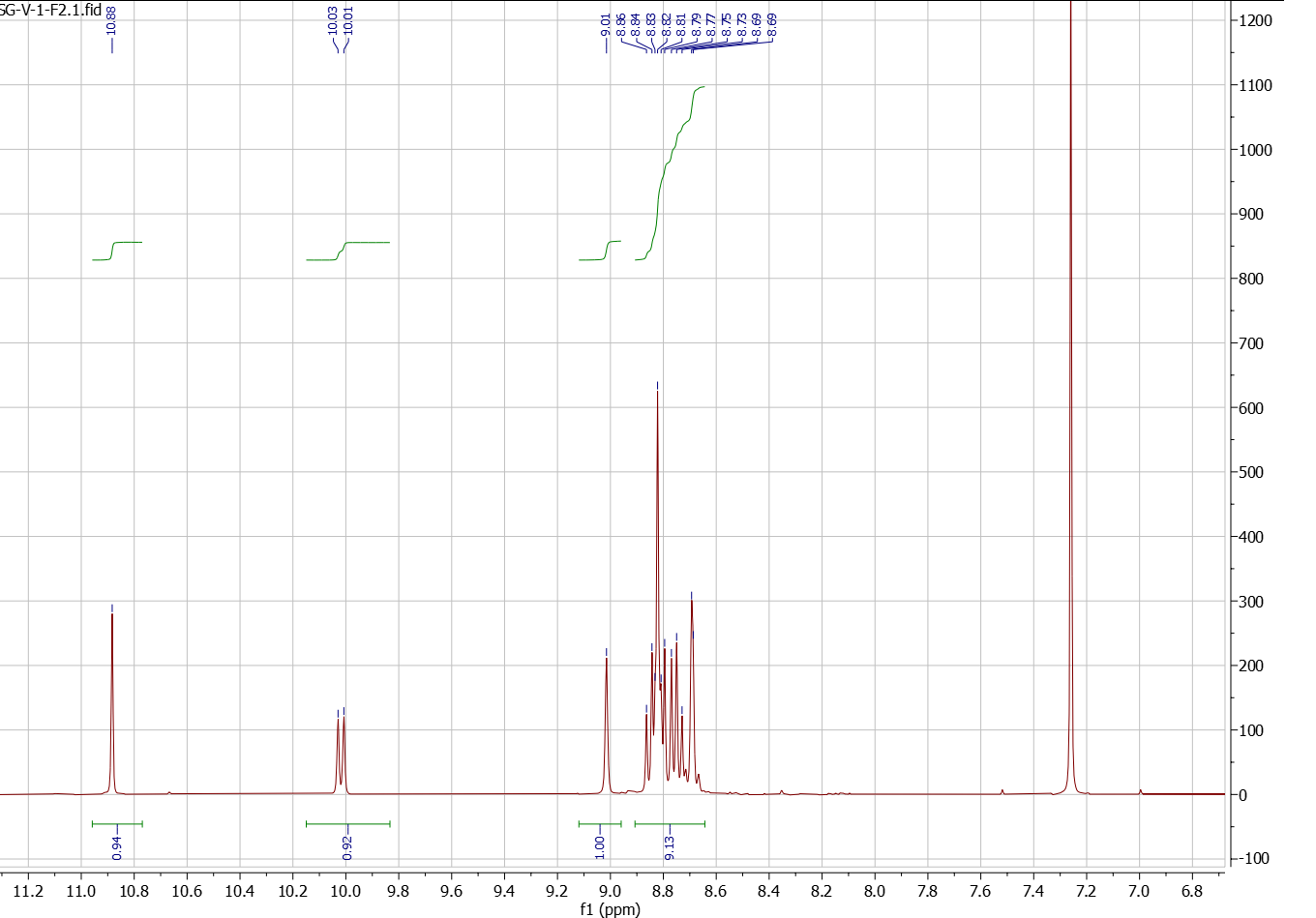}
    \includegraphics[width=0.8\linewidth]{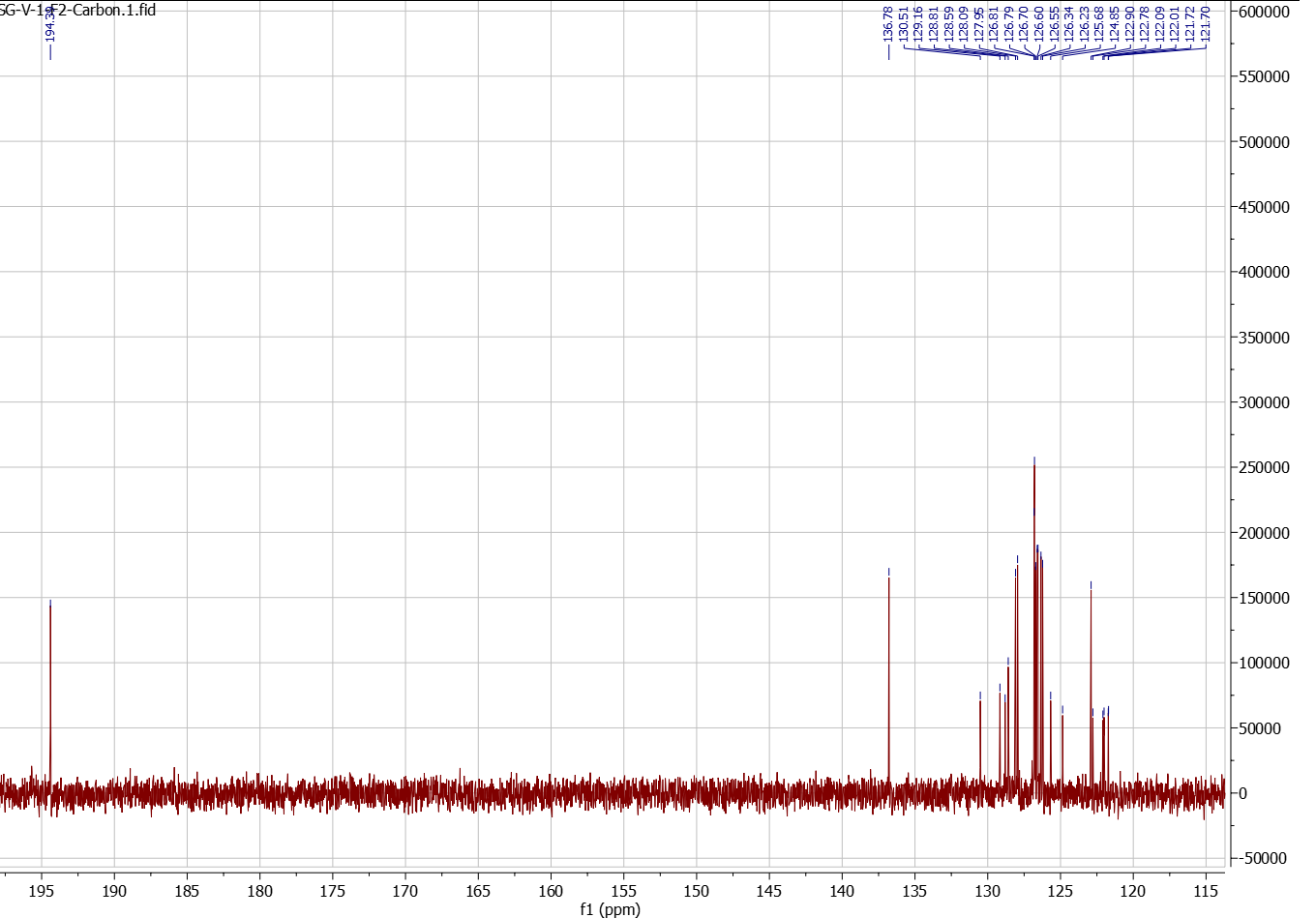}
    \caption{$^1$H NMR (top) and $^{13}$C NMR (bottom) spectrum of formylcoronene.}
    \label{fig:1HNMR_formylcoronene}
\end{figure}

Troubleshooting:
\begin{enumerate}
    \item The reaction will release \ce{HCl} gas during heating. Caution is advised.
    \item The reaction did not show improvement in yield after doubling the equivalence of \ce{SnCl4} and \ce{MOMCl2}. However, most of the starting material is recoverable.
    \item The product and unreacted starting material are both poorly soluble in a wide range of organic solvents. As such, copious amounts (potentially exceeding 6\,L) of DCM are required to perform chromatography at the scale indicated above. It is also recommended that the chromatography is completed as quickly as possible since crystallized products can severely clog the column.
\end{enumerate}

\subsection{Synthesis of Formylcoronene-Oxime}

To a 350\,mL pressure vessel was added formylcoronene (1.31\,g, 4.00\,mmol, 1\,eq), hydroxylammonium chloride (1.12\,g, 16.0\,mmol, 4\,eq), powdered 3 Angstrom mol sieves (800\,mg), mesitylene (40\,mL) and triethylamine (2.2\,mL, 16\,mmol, 4\,eq) in the above order. The vessel was immediately sealed and heated at 150\,$^\circ$C overnight. The mixture was then diluted with 500\,mL tetrahydrofuran (THF) and filtered. The THF was removed via rotary evaporator, and 100\,mL hexanes were added to precipitate the remaining formylcoronene-oxime. The solid precipitate was then filtered and washed with hexanes and methanol. The resulting solid was dried under vacuum to give formylcoronene-oxime (1.24\,g, 3.61\,mmol, 90\,\% yield) as a brown granular powder. $^1$H NMR (400\,MHz, \ce{CDCl3}): $\delta$ 9.57 (d, $J$ = 8.9\,Hz, 1H), 9.46 (s, 1H), 9.17 (s, 1H), 9.02 - 8.87 (m, 9H), 7.59 (s, 1H).\\

\begin{figure}
    \centering
    \includegraphics[width=0.8\linewidth]{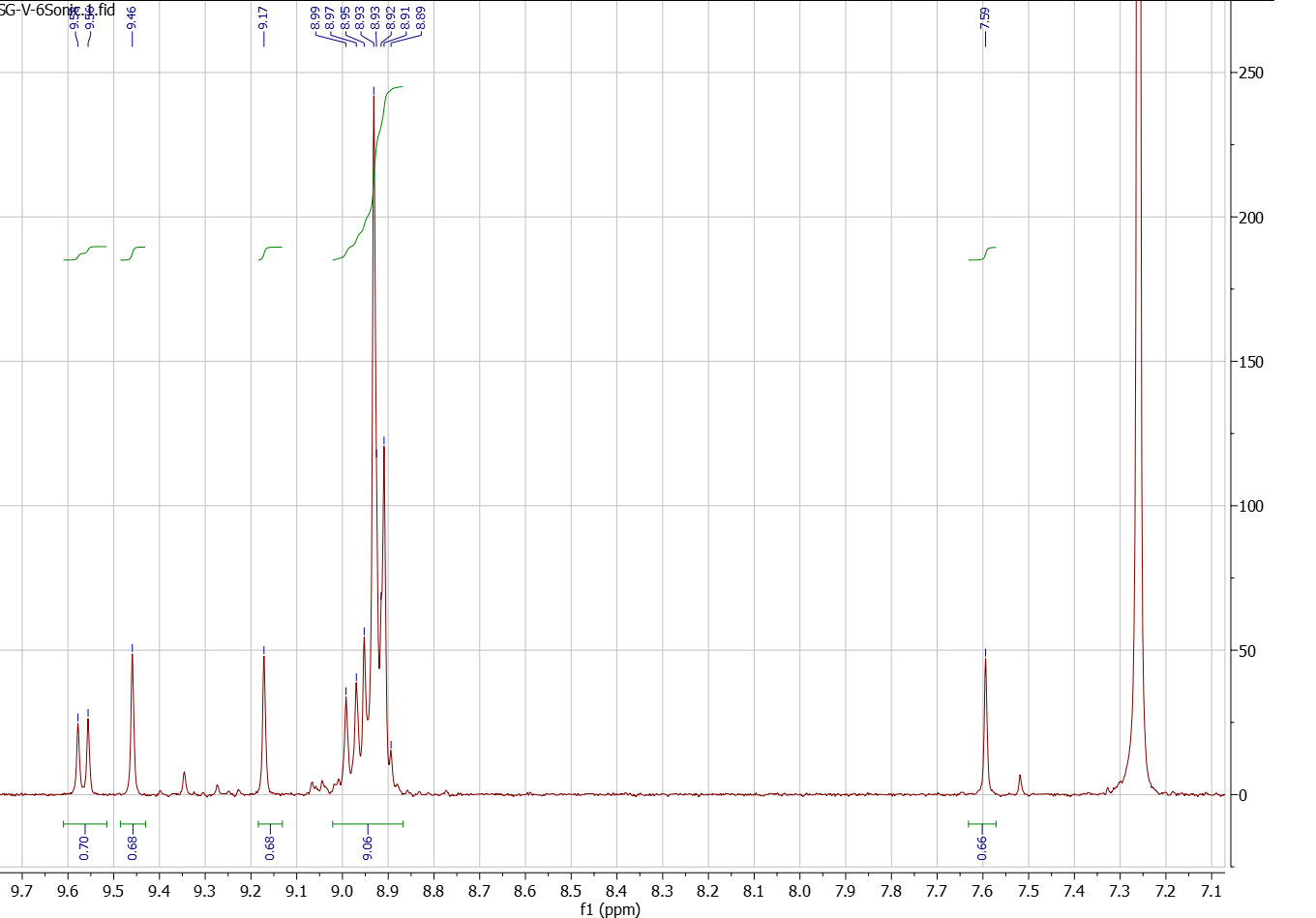}
    \caption{$^1$H NMR spectrum of formylcoronene-oxime.}
    \label{fig:1HNMR_formylcoronene-oxime}
\end{figure}

Troubleshooting:
\begin{enumerate}
    \item The reaction mixture is heterogeneous and can stick on the walls of the vessel if stirred too violently, thus leading to incomplete reaction. Slow stirring is preferred.
\end{enumerate}

\subsection{Synthesis of Cyanocoronene}

Cyanocoronene was prepared according to a modified literature procedure~\citep{Hyodo:2017:3005}: To an oven-dried round bottom flask under \ce{N2} was added formylcoronene-oxime (1.37\,g, 4\,mmol), 4-dimethylaminopyridine (DMAP) (49\,mg, 0.4\,mmol), DCM (20\,mL), triethylamine (NEt$_3$, 0.61\,mL, 4.4\,mmol) and \ce{PhSO3Cl} (0.51\,mL, 4\,mmol). The reaction was stirred overnight. Once the oxime was fully consumed via $^1$H NMR analysis, \ce{PhSO3H} (632\,mg, 4\,mmol) was dissolved in DCM (4\,mL) and injected into the reaction mixture. After stirring overnight, the reaction was quenched with NEt$_3$ and filtered through a silica pad with DCM as the eluent to obtain the product mixed with around 20\,\% co-eluting formylcoronene. The solvents were removed under reduced pressure to obtain a yellow solid (920\,mg). From the ratio of formylcoronene and cyanocoronene, the yield of cyanocoronene was calculated to be 2.35\,mmol (59\,\%). $^1$H NMR of the mixture (400\,MHz, \ce{CDCl3}): $\delta$ 10.98 (s, 0.2H, formylcoronene), 10.16 (d, $J$ = 8.8\,Hz, 0.2H, formylcoronene), 9.20 (s, 0.2H, formylcoronene), 9.13 – 9.05 (m, 2H, cyanocoronene), 8.98 - 8.79 (m, 12H, overlap), 8.72 (d, $J$ = 8.5 Hz, 1H, cyanocoronene).\\

\begin{figure}
    \centering
    \includegraphics[width=0.8\linewidth]{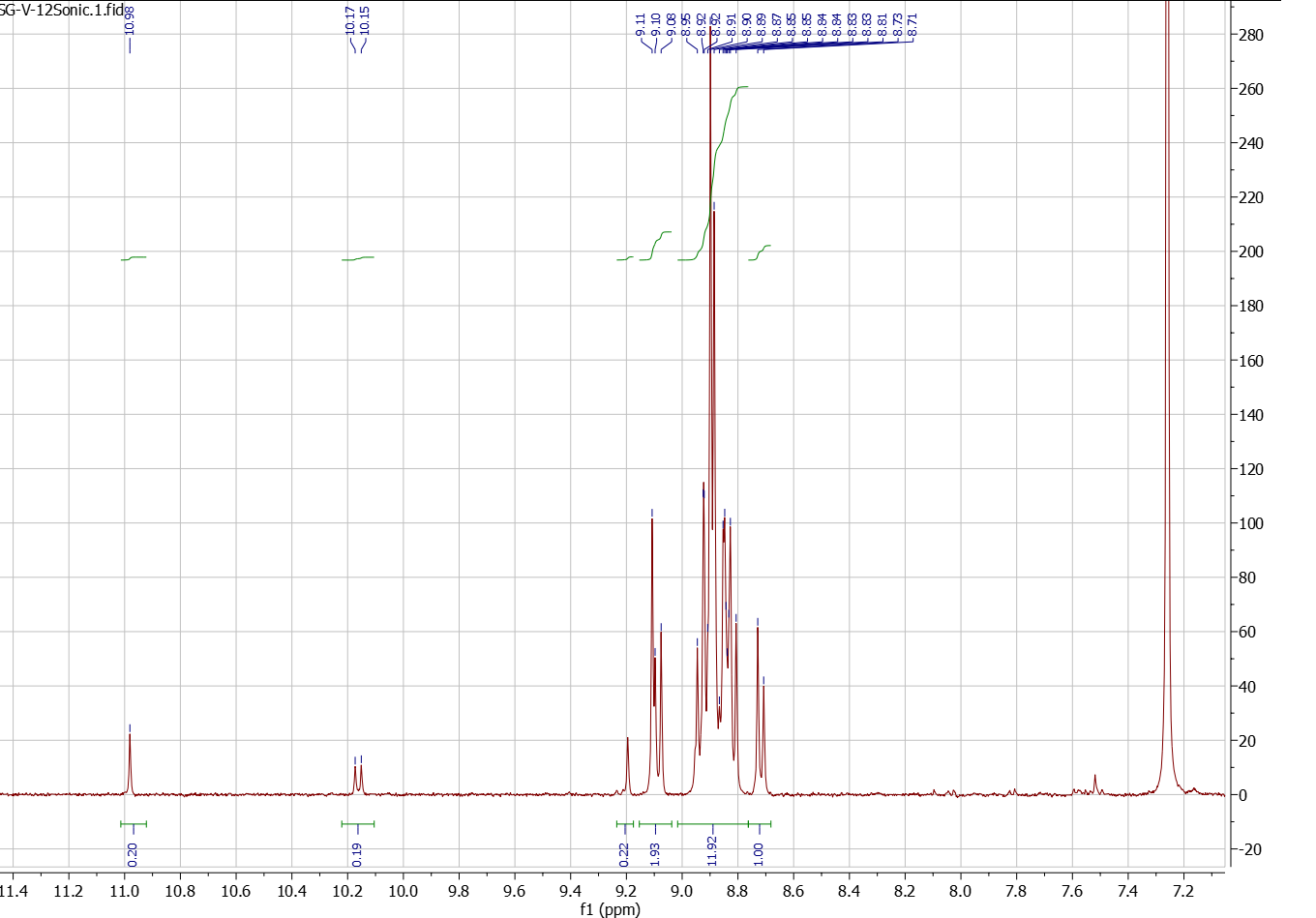}
    \caption{$^1$H NMR spectrum of cyanocoronene.}
    \label{fig:1HNMR_cyanocoronene}
\end{figure}

Troubleshooting:
\begin{enumerate}
    \item Direct conversion of formylcoronene to cyanocoronene according to literature protocol~\citep{Hyodo:2017:3005} is difficult due to the limited reactivity of formylcoronene at room temperature. Noticeable conversion is only reliably achieved in refluxing toluene, chlorobenzene, or mesitylene. However, the acidic reaction conditions combined with high temperatures frequently resulted in significant degradation of the starting material and severely decreased yields.
    \item The \ce{O}-sulfonyl oxime intermediate is sensitive to hydrolysis. Its isolation is not recommended.
    \item The reaction mixture is heterogeneous and can stick on the walls of the vessel if stirred too violently, thus leading to incomplete reaction. Slow stirring is preferred.
    \item Copious amounts (${\sim}$8\,L of DCM) were required to complete the chromatography due to the low solubility of the product. Precipitation-induced clogging can be mitigated by gently disturbing the top of the column.
\end{enumerate}

\clearpage

\section{Benchmark of Level of Theory}
\label{app:benchmark}

To benchmark the level of theory employed in this work, B3LYP/aug-cc-pVTZ, we provide in Table~\ref{tab:benchmark_b3lyp} theoretically computed rotational constants and quartic centrifugal distortion constants derived from the harmonic force field for the three cyanopyrene isomers which we previously studied using laboratory rotational spectroscopy~\citep{Wenzel:2024:810,Wenzel:2025:262}. The mean absolute percentage error (MAPE) comparing theoretical to experimental values for all three cyanopyrene isomers was 0.48\,\% and we therefore concluded that the B3LYP/aug-cc-pVTZ combination represents well the spectroscopic constants calculated for PAHs of this size.

\begin{table}[htb!]
    \centering
    \footnotesize
    \caption{Rotational constants of three cyanopyrene isomers in the A-reduced I$^r$ representation. Experimental values were taken from \citet{Wenzel:2024:810,Wenzel:2025:262}, theoretical values were calculated using the B3LYP/aug-cc-pVTZ level of theory, and the results of the combined theoretical and experimental fit are reported.}    
    \begin{tabular}{l r r r }
    
    \toprule
    & \multicolumn{3}{c}{\textbf{1-cyanopyrene}} \\
    Parameter               &  Theoretical   &   Experimental\tablenotemark{\footnotesize{a,b}}&  Theoretical+Experimental\tablenotemark{\footnotesize{a,c}}\\
    \midrule
    $A$ (MHz)               &  850.141 &  843.140191(128)  & 843.141827(128) \\
    $B$ (MHz)               &  372.931 &   372.500175(56) & 372.500183(56)  \\
    $C$ (MHz)               &  259.219 &   258.4249175(164) & 258.4248913(164) \\
    $\Delta_J$ (Hz)         &  1.977   &    2.240(80)  & 2.153(80))\\
    $\Delta_{JK}$ (Hz)      &  -5.556  &  -5.52(101)  & -5.88(101)  \\
    $\Delta_K$ (Hz)         &  20.310   &  [0]  & [20.310] \\
    $\delta_J$ (Hz)        &  0.724   & 0.826(40) & 0.795(40) \\ 
    $\delta_K$ (Hz)        &  3.059  & 5.26(74) & 4.20(74) \\
    \midrule\\
    & \multicolumn{3}{c}{\textbf{2-cyanopyrene}} \\
    Parameter               &  Theoretical   &   Experimental\tablenotemark{\footnotesize{a,b}}&  Theoretical+Experimental\tablenotemark{\footnotesize{a,c}}\\
    \midrule
    $A$ (MHz) & 1015.239 & 1009.19382(60) & 1009.19356(39) \\
    $B$ (MHz)               & 314.479 & 313.1345299(202) & 313.1345270(195) \\
    $C$ (MHz)               & 240.105 & 239.0427225(184) &239.0427270(167) \\
    $\Delta_J$ (Hz)         & 0.706 &  0.7008(109) & 0.7023(106) \\
    $\Delta_{JK}$ (Hz)      & 5.375 &  5.814(94) & 5.818(94)\\
    $\Delta_K$ (Hz)         & 8.785 & 15.3(114) & [8.785] \\
    $\delta_J$ (Hz)         & 0.184 & 0.1759(60)  & 0.1766(59) \\ 
    $\delta_K$ (Hz)         & 4.510 & 4.22(39) & 4.12(35)\\
    \midrule \\
     & \multicolumn{3}{c}{\textbf{4-cyanopyrene}} \\
    Parameter               &  Theoretical   &   Experimental\tablenotemark{\footnotesize{a,b}}&  Theoretical+Experimental\tablenotemark{\footnotesize{a,c}}\\
    \midrule
    $A$ (MHz) & 652.155 &  651.383034(69) & 651.382955(64)\\
    $B$ (MHz) & 456.670 &  453.731352(45) & 453.731444(41)\\
    $C$ (MHz) & 268.590 & 267.5078168(224) & 267.5078004(221)\\
    $\Delta_J$ (Hz) & 1.893 & 1.780(98) & 2.026(69)\\
    $\Delta_{JK}$ (Hz) & -1.553 & 0.43(58) &  [-1.553]\\
    $\Delta_K$ (Hz) & 14.821 &  10.89(73) & 12.61(53)\\
    $\delta_J$ (Hz) & 0.756 & 0.677(49) & 0.803(35)\\ 
    $\delta_K$ (Hz) & 1.971 &  2.56(33) & 2.40(33) \\
    \bottomrule 
    \end{tabular}
    {\footnotesize\tablenotetext{a}{Values in parentheses are 1$\,\sigma$ uncertainties in units of the last digit.}
    \tablenotetext{b}{Values from \citet{Wenzel:2024:810,Wenzel:2025:262}. The value in brackets was not determinable and fixed to zero.}
    \tablenotetext{c}{Values in brackets possessed large uncertainties and were fixed to their respective theoretical constants.}}
    \label{tab:benchmark_b3lyp}
\end{table}
 
\clearpage
\section{Measured Lines of Cyanocoronene}

\begin{figure}
    \centering
    \includegraphics[width=0.875\linewidth]{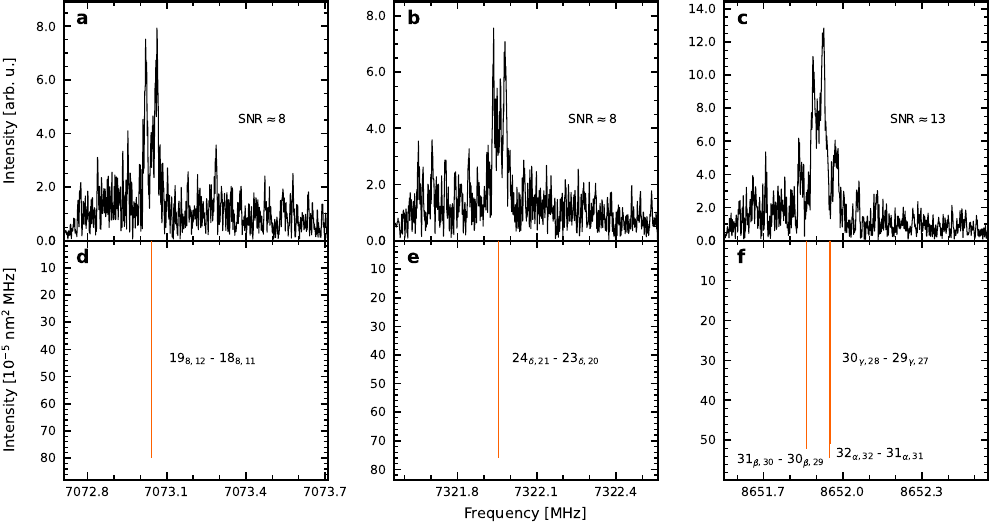}
    \caption{Example transitions of cyanocoronene recorded with the cavity-enhanced FTMW spectrometer. (a -- c) Three example spectra that were recorded by averaging approximately 4500 to 10000 gas pulses acquired at a 5\,Hz repetition rate and a 1\,$\upmu$s excitation pulse. 
    The lines are split due to Doppler doubling induced by the co-axial cavity--supersonic expansion geometry of the instrument. We also report the SNR of the recorded lines. (d -- f) Orange lines indicate the fitted rest frequencies. Corresponding quantum numbers of the transitions, $J^\prime_{K_a^\prime, K_c^\prime} - J^{\prime\prime}_{K_a^{\prime\prime}, K_c^{\prime\prime}}$, are labelled. Closely overlapping $K_a$-components are labelled with $\alpha = \{0,1\}, \beta = \{1,2\}, \gamma = \{2,3\}, \delta = \{3,4\}$.}
    \label{fig:cavity_lines}
\end{figure}

The experimentally measured lines of cyanocoronene are presented in Table~\ref{tab:transitions}. Examples of experimentally recorded transitions using the cavity-enhanced FTMW spectrometer are depicted in Fig.~\ref{fig:cavity_lines}.

\startlongtable
\begin{deluxetable}{cc}
    \tablewidth{8in}
    \tablecaption{Experimentally measured transitions of cyanocoronene between 6788 and 10520\,MHz.\label{tab:transitions}}
    \tablehead{
    \colhead{Transition} & \colhead{Frequency\tablenotemark{a}} \\
     \colhead{($J_{K_a^{\prime},K_c^{\prime}}^{\prime} - J_{K_a^{\prime\prime},K_c^{\prime\prime}}^{\prime\prime}$)} & \colhead{(MHz)}
     }
    \startdata
    $24_{1,23} - 23_{1,22}$ & 6788.4098 \\
    $24_{2,23} - 23_{2,22}$ & 6788.4098 \\
    $25_{0,25} - 24_{0,24}$ & 6788.4443 \\
    $25_{1,25} - 24_{1,24}$ & 6788.4443 \\
    $23_{2,21} - 22_{2,20}$ & 6788.7145 \\
    $23_{3,21} - 22_{3,20}$ & 6788.7145 \\
    $22_{4,19} - 21_{4,18}$ & 6789.9187 \\
    $22_{3,19} - 21_{3,18}$ & 6789.9187 \\
    $19_{7,13} - 18_{7,12}$ & 6815.8858 \\
    $18_{7,11} - 17_{7,10}$ & 7005.2518 \\
    $25_{1,24} - 24_{1,23}$ & 7054.6106 \\
    $25_{2,24} - 24_{2,23}$ & 7054.6106 \\
    $26_{0,26} - 25_{0,25}$ & 7054.6617 \\
    $26_{1,26} - 25_{1,25}$ & 7054.6617 \\
    $24_{2,22} - 23_{2,21}$ & 7054.8759 \\
    $24_{3,22} - 23_{3,21}$ & 7054.8759 \\
    $23_{3,20} - 22_{3,19}$ & 7055.9209 \\
    $23_{4,20} - 22_{4,19}$ & 7055.9209 \\
    $22_{5,18} - 21_{5,17}$ & 7058.7857 \\
    $22_{4,18} - 21_{4,17}$ & 7058.7857 \\
    $19_{8,12} - 18_{8,11}$ & 7073.0418 \\
    $20_{7,14} - 19_{7,13}$ & 7080.6353 \\
    $20_{6,14} - 19_{6,13}$ & 7088.6004 \\
    $18_{8,10} - 17_{8,9}$ & 7291.4710 \\
    $26_{1,25} - 25_{1,24}$ & 7320.8179 \\
    $26_{2,25} - 25_{2,24}$ & 7320.8179 \\
    $27_{0,27} - 26_{0,26}$ & 7320.8780 \\
    $27_{1,27} - 26_{1,26}$ & 7320.8780 \\
    $25_{2,23} - 24_{2,22}$ & 7321.0420 \\
    $25_{3,23} - 24_{3,22}$ & 7321.0420 \\
    $24_{3,21} - 23_{3,20}$ & 7321.9574 \\
    $24_{4,21} - 23_{4,20}$ & 7321.9574 \\
    $23_{5,19} - 22_{5,18}$ & 7324.4393 \\
    $23_{4,19} - 22_{4,18}$ & 7324.4393 \\
    $27_{1,26} - 26_{1,25}$ & 7587.0283 \\
    $27_{2,26} - 26_{2,25}$ & 7587.0283 \\
    $28_{0,28} - 27_{0,27}$ & 7587.0875 \\
    $28_{1,28} - 27_{1,27}$ & 7587.0875 \\
    $22_{10,13} - 21_{10,12}$ & 8377.6038 \\
    $30_{1,29} - 29_{1,28}$ & 8385.6505 \\
    $30_{2,29} - 29_{2,28}$ & 8385.6505 \\
    $31_{0,31} - 30_{0,30}$ & 8385.7347 \\
    $31_{1,31} - 30_{1,30}$ & 8385.7347 \\
    $29_{2,27} - 28_{2,26}$ & 8385.7636 \\
    $29_{3,27} - 28_{3,26}$ & 8385.7636 \\
    $26_{6,21} - 25_{6,20}$ & 8391.1806 \\
    $26_{5,21} - 25_{5,20}$ & 8391.1806 \\
    $31_{1,30} - 30_{1,29}$ & 8651.8594 \\
    $31_{2,30} - 30_{2,29}$ & 8651.8594 \\
    $32_{0,32} - 31_{0,31}$ & 8651.9466 \\
    $32_{1,32} - 31_{1,31}$ & 8651.9466 \\
    $30_{2,28} - 29_{2,27}$ & 8651.9537 \\
    $30_{3,28} - 29_{3,27}$ & 8651.9537 \\
    $29_{3,26} - 28_{3,25}$ & 8652.4539 \\
    $29_{4,26} - 28_{4,25}$ & 8652.4539 \\
    $28_{4,24} - 27_{4,23}$ & 8653.7843 \\
    $28_{5,24} - 27_{5,23}$ & 8653.7843 \\
    $27_{6,22} - 26_{6,21}$ & 8656.7527 \\
    $27_{5,22} - 26_{5,21}$ & 8656.7527 \\
    $37_{2,35} - 36_{2,34}$ & 10515.3437 \\
    $37_{3,35} - 36_{3,34}$ & 10515.3437 \\
    $38_{1,37} - 37_{1,36}$ & 10515.3437 \\
    $38_{2,37} - 37_{2,36}$ & 10515.3437 \\
    $39_{0,39} - 38_{0,38}$ & 10515.4529 \\
    $39_{1,39} - 38_{1,38}$ & 10515.4529 \\
    $35_{4,31} - 34_{4,30}$ & 10516.2356 \\
    $35_{5,31} - 34_{5,30}$ & 10516.2356 \\
    $34_{5,29} - 33_{5,28}$ & 10517.6703 \\
    $34_{6,29} - 33_{6,28}$ & 10517.6703 \\
    $33_{6,27} - 32_{6,26}$ & 10520.4313 \\
    $33_{7,27} - 32_{7,26}$ & 10520.4313 \\
    \enddata
    \tablenotetext{a}{Experimental uncertainties are 2\,kHz.}
\end{deluxetable}

\section{Partition Function for Cyanocoronene}

The partition function calculated by \textsc{SPCAT} for cyanocoronene, excluding $^{14}$N-quadrupole hyperfine splitting, is reported in Table~\ref{tab:qrot}.

\begin{table}[htb!]
    \centering
    \caption{Partition function for cyanocoronene which was calculated by \textsc{SPCAT} for $J_\mathrm{max} = 400$ and $K_\mathrm{max} = 355$ and used in the MCMC analysis.}
    \begin{tabular}{rrr}
    \toprule
        {Temperature [K]} & & {Partition function} \\
    \hline
1.0 && 1703.7230 \\
2.0 &&  4813.7245 \\
3.0 &&  8840.2413 \\
4.0 &&  13608.0333 \\
5.0 &&  19015.7897 \\
6.0 &&  24995.1437 \\
7.0 &&  31495.9011 \\
8.0 &&  38479.1687 \\
9.375 &&  48812.4929 \\
18.75 &&  138047.6140 \\
37.5 &&  390439.4924 \\
75.0 &&  1104317.3292 \\
150.0 &&  3120078.8657 \\
225.0 &&  5686759.1241 \\
300.0 &&  8589526.6725 \\
400.0 &&  12678742.2060 \\
500.0 &&  16744897.1786 \\
        \bottomrule
    \end{tabular}
    \label{tab:qrot}
\end{table}

\clearpage
\section{MCMC Analysis of Cyanocoronene in TMC-1}

\begin{figure*}
\includegraphics[width=\textwidth]{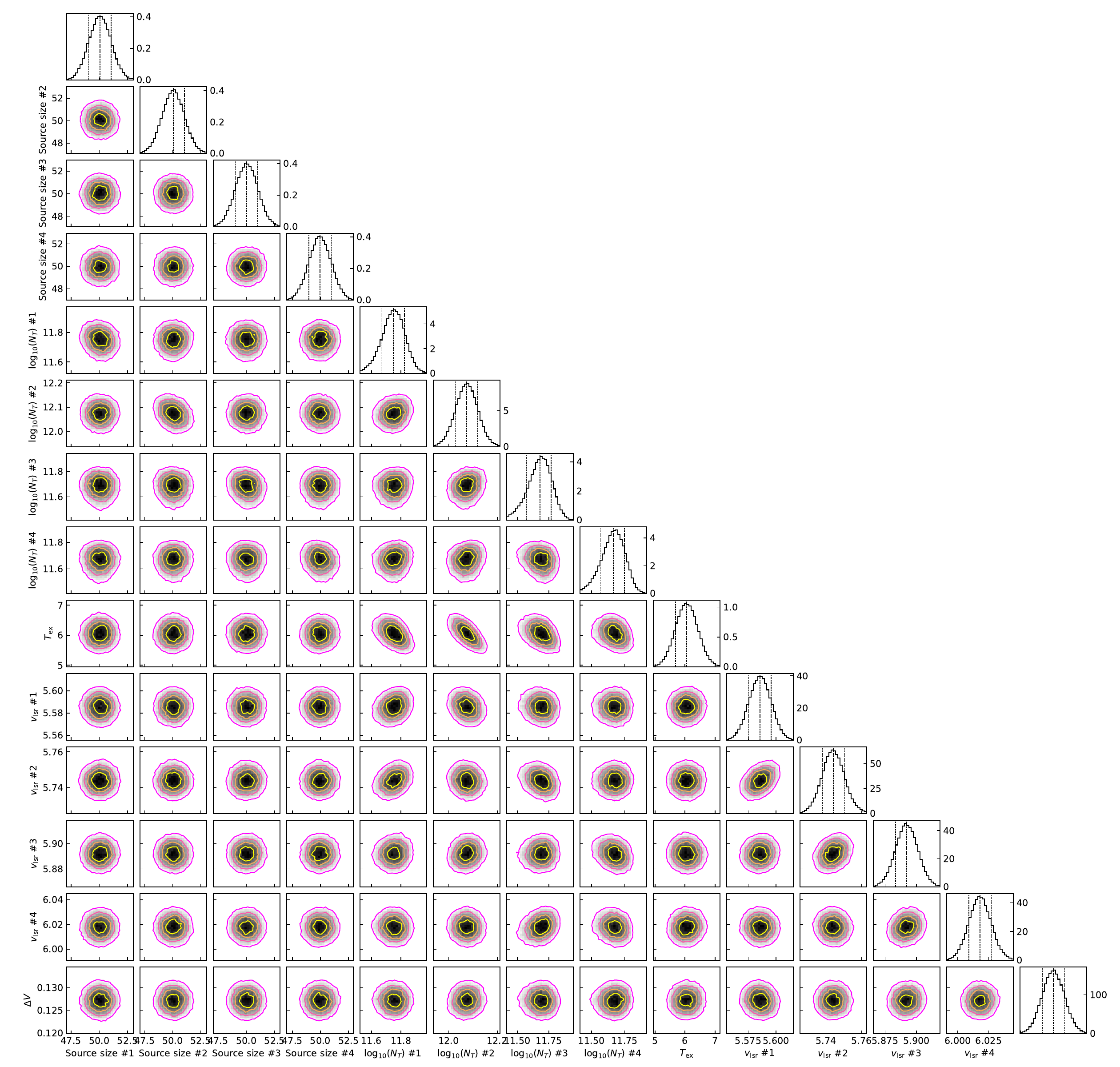}
    \caption{Parameter covariances and marginalized posterior distributions for the MCMC fit of cyanocoronene. 16$^\mathrm{th}$, 50$^\mathrm{th}$, and 84$^\mathrm{th}$ confidence intervals (corresponding to $\pm1\,\sigma$ for a Gaussian posterior distribution) are depicted as vertical lines on the diagonal plots.}
    \label{fig:cornerplot}
\end{figure*}

The priors used for the MCMC analysis are reported in Table~\ref{tab:MCMCpriors} and the resulting marginalized posterior distributions are listed and depicted in Table~\ref{tab:mcmc} and Fig.~\ref{fig:cornerplot}, respectively. Note that we ran our MCMC analysis on 7768 transitions that have an uncertainty of lower than 5\,kHz in the cyanocoronene catalog. The priors for velocity, $v_\mathrm{LSR}$, source size, and linewidth, $\Delta V$, were heavily constrained with Gaussian distributions, $\mathcal{N}(\mu,\sigma^2)$, with mean, $\mu$ and variance, $\sigma^2$, informed from prior observations of PAHs~\citep{McGuire:2021:1265,Wenzel:2024:810,Wenzel:2025:262}, while the column density, $N_\mathrm{T}$, and excitation temperature, $T_\mathrm{ex}$, were allowed to vary freely using a uniform (unweighted) distribution, $\mathcal{U}\{a,b\}$, between $a$ and $b$.

\begin{table}[htb!]
    \caption{Priors used for the MCMC analysis. A Gaussian distribution, $\mathcal{N}(\mu,\sigma^2)$, with mean $\mu$ and variance $\sigma^2$, was used for the source size, velocity, $v_\mathrm{LSR}$, and linewidth, $\Delta V$, and uniform (unweighted) distribution, $\mathcal{U}\{a,b\}$, between $a$ and $b$ was used for the column density, $N_\mathrm{T}$, and excitation temperature, $T_\mathrm{ex}$.}
    \centering
    \begin{tabular}{cccccccccc}
    \\
    \toprule
      Component &  $v_\mathrm{LSR}$	&	Size	&	$\mathrm{log_{10}}(N_\mathrm{T})$	&	$T_\mathrm{ex}$	&	$\Delta V$	\\
	No. & ($\mathrm{km\,s^{-1}}$) &($^{\prime\prime}$)	&	($\mathrm{cm}^{-2}$)	&	($\mathrm{K}$)	&	($\mathrm{km\,s^{-1}}$)\\
	\midrule
	1 & {$\mathcal{N}(5.575,0.01)$} & \multirow{4}{*}{$\mathcal{N}(50,1)$}	 &  	 \multirow{4}{*}
 {$\mathcal{U}\{\mathrm{a,b}\}$} &
 \multirow{4}{*}{$\mathcal{U}\{\mathrm{a,b}\}$}	 & 	 \multirow{4}{*}{$\mathcal{N}(0.125,0.005)$}\\
	2 &	{$\mathcal{N}(5.767,0.01)$} &	 & 	 & 	 & 	 \\
        3 & {$\mathcal{N}(5.892,0.01)$} &	 & 	 &   & 	 \\
	4 &	{$\mathcal{N}(6.018,0.01)$} &  &	 & 	 & 	 \\
 \midrule
        Min & $0.0$ & $1$ & $10.0$ & $3.0$ & $0.1$ \\
        Max & $10.0$ & $100$ & $13.0$ & $15.0$ & $0.3$ \\
	\bottomrule
\\
    \end{tabular}\\
    \label{tab:MCMCpriors}
\end{table}

\begin{table}[hbt!]
    \centering
    \caption{Summary statistics of the marginalized posterior probability distributions from the MCMC analysis for cyanocoronene. Priors used are reported in Table~\ref{tab:MCMCpriors}. The uncertainties are represented by the 16$^\mathrm{th}$ and 84$^\mathrm{th}$ percentile, also known as the $68\,\%$ confidence interval, which corresponds to 1$\,\sigma$ for a Gaussian distribution. The total column density and its uncertainty are derived by marginalizing over the posterior distributions for the column densities of each individual component and reporting the 50$^\mathrm{th}$, 16$^\mathrm{th}$, and 84$^\mathrm{th}$ percentiles. Source sizes were artificially over-constrained (see \citealt{Wenzel:2024:810}), and hence, we do not report their uncertainties.}
    \begin{tabular}{ccccccccccc}
    \\
    \toprule
      Component & $v_\mathrm{LSR}$	&	Size	&	$N_\mathrm{T}$	&	$T_\mathrm{ex}$	&	$\Delta V$	\\
	No. & {($\mathrm{km\,s^{-1}}$)} & ($^{\prime\prime}$)	&	($10^{12}\,\mathrm{cm}^{-2}$)	&	($\mathrm{K}$)	&	($\mathrm{km\,s^{-1}}$)\\
    \midrule
	1 &	$5.586^{+0.010}_{-0.010}$	 & 	$50$	 & 	$0.56^{+0.11}_{-0.10}$	 & 	 \multirow{4}{*}{$6.05^{+0.38}_{-0.37}$}	 & 	 \multirow{4}{*}{$0.127^{+0.002}_{-0.002}$}\\
	2 &	$5.744^{+0.006}_{-0.006}$	 & 	$50$	 & 	$1.19^{+0.13}_{-0.12}$	 & 	 & 	 \\		3 & $5.892^{+0.009}_{-0.009}$	 & 	$50$	 & 	$0.48^{+0.11}_{-0.10}$	 & 	 & 	 \\
	4 &	$6.018^{+0.009}_{-0.009}$	 & 	$49$	 & 	$0.46^{+0.10}_{-0.09}$	 & 	 & 	 \\
	\midrule
		\multicolumn{6}{c}{$N_\mathrm{T}(\ce{C24H11CN}) = N_\mathrm{T}(\mathrm{summed}) = \;2.69^{+0.26}_{-0.23}\times 10^{12}\,\mathrm{cm}^{-2}$}\\
  \bottomrule
    \end{tabular}
    \label{tab:mcmc}
\end{table}

\clearpage
\section{Column Densities of Cyclic Hydrocarbons in TMC-1}

The data compiled in Fig.~\ref{fig:CDcomp} is listed in Table~\ref{tab:CDcomp}.
\begin{table}[!htb]
    \centering
    \caption{Column densities of carbon-bearing cycles observed in TMC-1 as presented in Fig.~\ref{fig:CDcomp}.}    
    \begin{tabular}{c c c c}
    \\
    \toprule
    Molecule      & $N_\mathrm{C}$ & Column Density & Reference\\
                 &   &    ($10^{12}\,\mathrm{cm}^{-2}$)  \\
     \midrule
    \ce{C5H6}       & 5 & 12$^{+3}_{-3}$  &~\citet{Cernicharo:2021:L15} \\
    \ce{C6H4}       & 6 & 0.5$^{+0.1}_{-0.1}$  &~\citet{Cernicharo:2021:L9} \\
    1-\ce{C5H5CN}   & 6 & 0.827$^{+0.090}_{-0.100}$ &~\citet{Lee:2021:L2}         \\
    2-\ce{C5H5CN}   & 6 & 0.189$^{+0.018}_{-0.015}$  &~\citet{Lee:2021:L2}         \\
    1-\ce{C5H5CCH}       & 7 & 1.4$^{+0.2}_{-0.2}$  &~\citet{Cernicharo:2021:L1FORCE} \\ 
    2-\ce{C5H5CCH}       & 7 & 2.0$^{+0.4}_{-0.4}$  &~\citet{Cernicharo:2021:L1FORCE} \\ 
    \ce{C5H4CCH2}    & 7 &  2.7$^{+0.3}_{-0.3}$ &~\citet{Cernicharo:2022:L9}\\    
    \ce{C6H5CN}    & 7 &  1.73$^{+0.85}_{-0.10}$ &~\citet{McGuire:2021:1265}\\
    \ce{C6H5CCH}       & 8 & 3.0$^{+0.5}_{-0.5}$  &~\citet{Loru:2023:A166} \\  
    \ce{C9H8}      & 9 &  9.04$^{+0.96}_{-0.96}$ &~\citet{Sita:2022:L12}                \\
    2-\ce{C9H7CN}   & 10 &   0.210$^{+0.060}_{-0.046}$       &~\citet{Sita:2022:L12}           \\
    1-\ce{C10H7CN} & 11 &  0.74$^{+0.33}_{-0.46}$  &~\citet{McGuire:2021:1265}\\
    2-\ce{C10H7CN} & 11 &  0.71$^{+0.45}_{-0.32}$ &\citet{McGuire:2021:1265} \\
    1-\ce{C12H7CN} & 13 & 0.95$^{+0.09}_{-0.09}$ &\citet{Cernicharo:2024:L13} \\
    5-\ce{C12H7CN} & 13 & 0.95$^{+0.09}_{-0.09}$  &\citet{Cernicharo:2024:L13} \\
    1-\ce{C16H9CN} & 17 & 1.52$^{+0.18}_{-0.16}$ &~\citet{Wenzel:2024:810}\\
    2-\ce{C16H9CN} & 17 & 0.84$^{+0.09}_{-0.09}$ &~\citet{Wenzel:2025:262}\\
    4-\ce{C16H9CN} & 17 & 1.33$^{+0.10}_{-0.09}$ &~\citet{Wenzel:2025:262}\\
    \ce{C16H10}\tablenotemark{\footnotesize{*}} & 16 & ${\sim}$30  &~\citet{Wenzel:2025:262} \\ 
    \ce{C24H11CN} & 25 & 2.69$^{+0.26}_{-0.23}$ & This work\\
    \ce{C24H12}\tablenotemark{\footnotesize{*}} & 24 & ${\sim}$20  & This work \\ 
    \bottomrule
    \end{tabular}
    \tablenotetext{*}{Estimated using the cyanopyrene and cyanocoronene column densities, calculated rate coefficients for \ce{CN} addition, and an \ce{CN}/\ce{H} ratio of 0.15 (see Appendix~\ref{app:mesmer} and \citealt{Wenzel:2025:262}).}
    \label{tab:CDcomp}
\end{table}

\section{Ab-initio quantum chemical calculations}
\label{app:mesmer}

Master equation calculations were performed in MESMER 7.1 ~\citep{Glowacki:2012:9545} on an ab-intio surface calculated at the DLPNO-CCSD/def2-TZVPP//$\omega$B97X-D4/def2-TZVPP level ~\citep{Chai:2008:084106,Weigend:2006:1057,Caldeweyher:2017:034112} in ORCA 5.0.4~\citep{Neese:2022:e1606}; the master equation calculations showed that below 100\,K with a gas density of $2 \times 10^{4}\,\mathrm{cm}^{-3}$ the reaction of \ce{CN} with coronene leads to the formation of cyanocoronene and an \ce{H} atom at the collision rate. The collision rate is predicted to be $k_\mathrm{col} = 5.6^{+5.6}_{-2.8} \times 10^{-10}\,\mathrm{cm^{3} s^{-1}}$ using classical capture theory in the manner described in \citet{
West:2019:134,Wenzel:2025:262}; with no prediction of the formation of isocyanocoronene. Using this collision rate and its relationship to the \ce{CN}/\ce{H} ratio on formation rate from \citet{Wenzel:2025:262}, \ce{CN}/\ce{H} = 0.15, leads to a ratio of coronene to cyanocoronene of 6.67 and a column density of coronene of $N(\ce{C24H12}) \approx 2 \times 10^{13}\,\mathrm{cm}^{-2}$.

\subsection{Computational Methodology}
Calculations of a potential energy surface for the formation of cyanocoronene were carried out in ORCA 5.0.4~\citep{Neese:2022:e1606}.  Initial structures for the adducts and separated reagents were optimized with the RI-BP86 DFT functional~\citep{Perdew:1986:8822,Becke:1988:3098,Lee:1988:785} using the def2-SVP basis set~\citep{Weigend:2005:3297} and including D3(BJ) empirical dispersion corrections~\citep{Grimme:2011:1456,Becke:2006:221101} (this method will be subsequently referred to as DFT-Cheap). The presence of barriers to the addition of \ce{CN} to coronene and the subsequent elimination of an H atom was evaluated by carrying out modified redundant scans of the forming or breaking bonds. In these scans, all coordinates except the bond being scanned were allowed to relax while the scanned coordinate was varied. The resultant maxima were used as input geometries for transition state optimizations and the long-range minima prior to \ce{CN} addition were optimized to a loose bound complex on the entrance channel.

Harmonic vibrational analysis was carried out to verify that these structures were indeed stationary points and the resulting frequencies were scaled by 0.9956 and harmonic zero point energies were scaled 1.0207 in the manner suggested by~\citet{Kesharwani:2015:1701}. Intrinsic reaction coordinate (IRC) scans were performed to verify that transition states linked reactants and products~\citep{Ishida:1977:2153,Neese:2022:e1606}. The structures found by this approach were re-optimized with the hybrid $\omega$B97X functional~\citep{Chai:2008:084106}  with the D4 empirical dispersion correction~\citep{Caldeweyher:2017:034112} and the def2-TZVPP triple zeta basis set~\citep{Weigend:2006:1057} (this method will subsequently be referred to as DFT-2). Harmonic vibrational frequencies were calculated on the DFT-2 structures and the results were scaled by 0.9533 and harmonic zero point energies were scaled 0.9779 in the manner suggested by ~\citet{Kesharwani:2015:1701}. 

Single point energy corrections were carried out using the domain pair local natural orbital coupled cluster method including single and double excitations DLPNO-CCSD, with tight thresholds for the PNOs~\citep{Pinski:2015:034108,Riplinger:2016:024109}. This method was used with two basis sets, firstly, using Ahlrichs triple zeta def2-TZVPP basis set and secondly with Dunning's augmented double zeta correlation consistent basis set aug-cc-pVDZ~\citep{Dunning:1989:1007a,Kendall:1992:6796,Davidson:1996:514}. These single-point energy corrections were calculated for both of the structures calculated with DFT-Cheap and DFT-2. For the reactants and products, additional single point energy calculations were carried out at the CCSD(T)/cc-pVDZ level. These results were combined with three MP2 calculations with three of Dunning's correlation consistent basis set series cc-pVDZ, cc-pVTZ, cc-pVQZ~\citep{Dunning:1989:1007a,Davidson:1996:514}; taken from the correlation consistent basis set repository, ccREPO~\citep{ccREPO-2016}. These results were used with equation~\ref{eq:EP3} to calculate an EP3 approximation of the CCSD(T) complete basis set limit ~\citep{Jurecka:2006:1985,Liakos:2012:4801}. 

\begin{equation}\label{eq:EP3}
 \textrm{CCSD(T)/CBS} \approx \textrm{EP3} = E_{\textrm{HFCBS}} + E_{\textrm{MP2CBS}} + E_{\textrm{CCSD(T)small}} - E_{\textrm{MP2small}} 
 \end{equation}
 
\begin{figure*}[bt]
    \centering
    \includegraphics[width=0.90\textwidth]{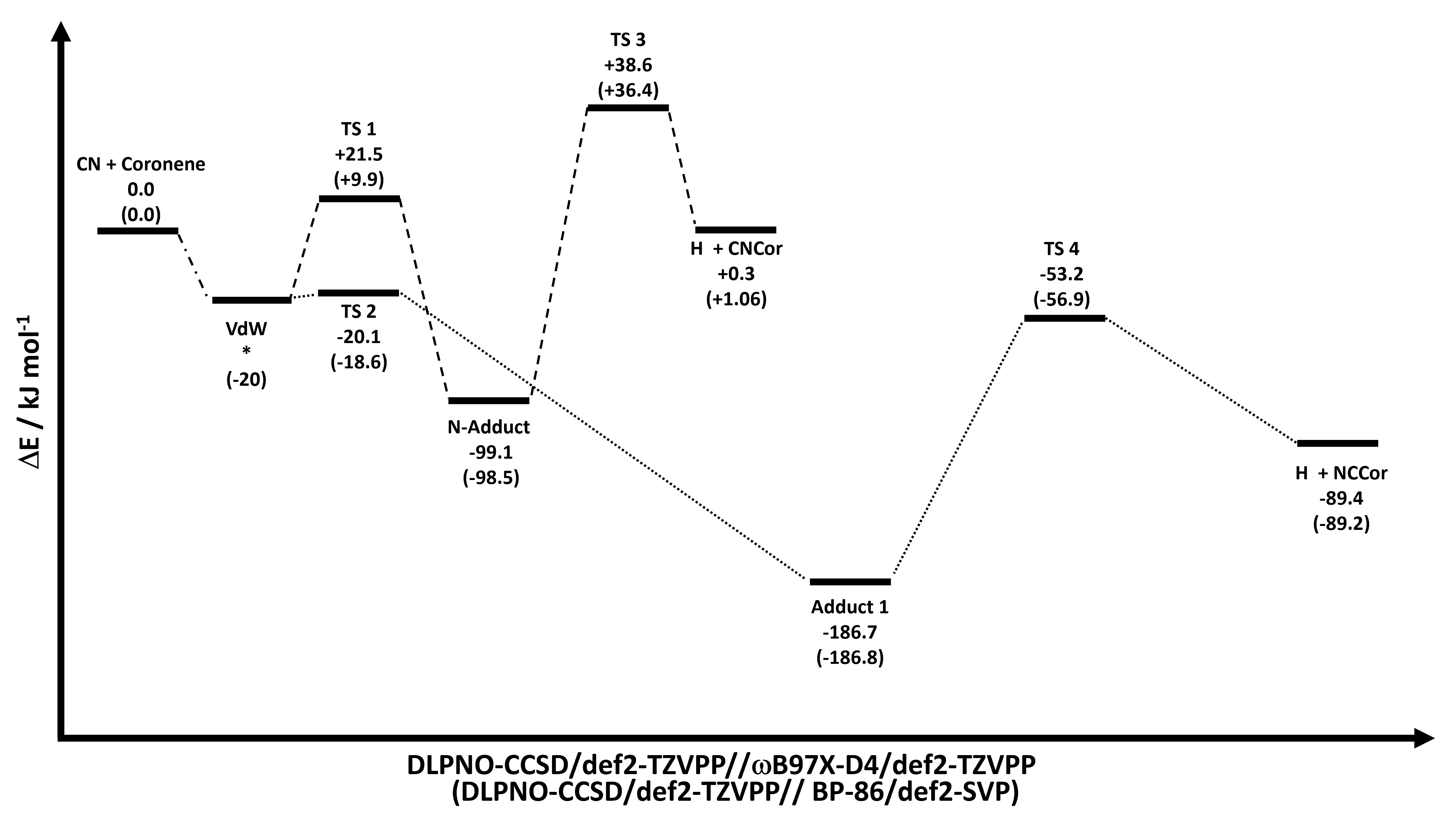}\\
    \includegraphics[width=0.90\textwidth]{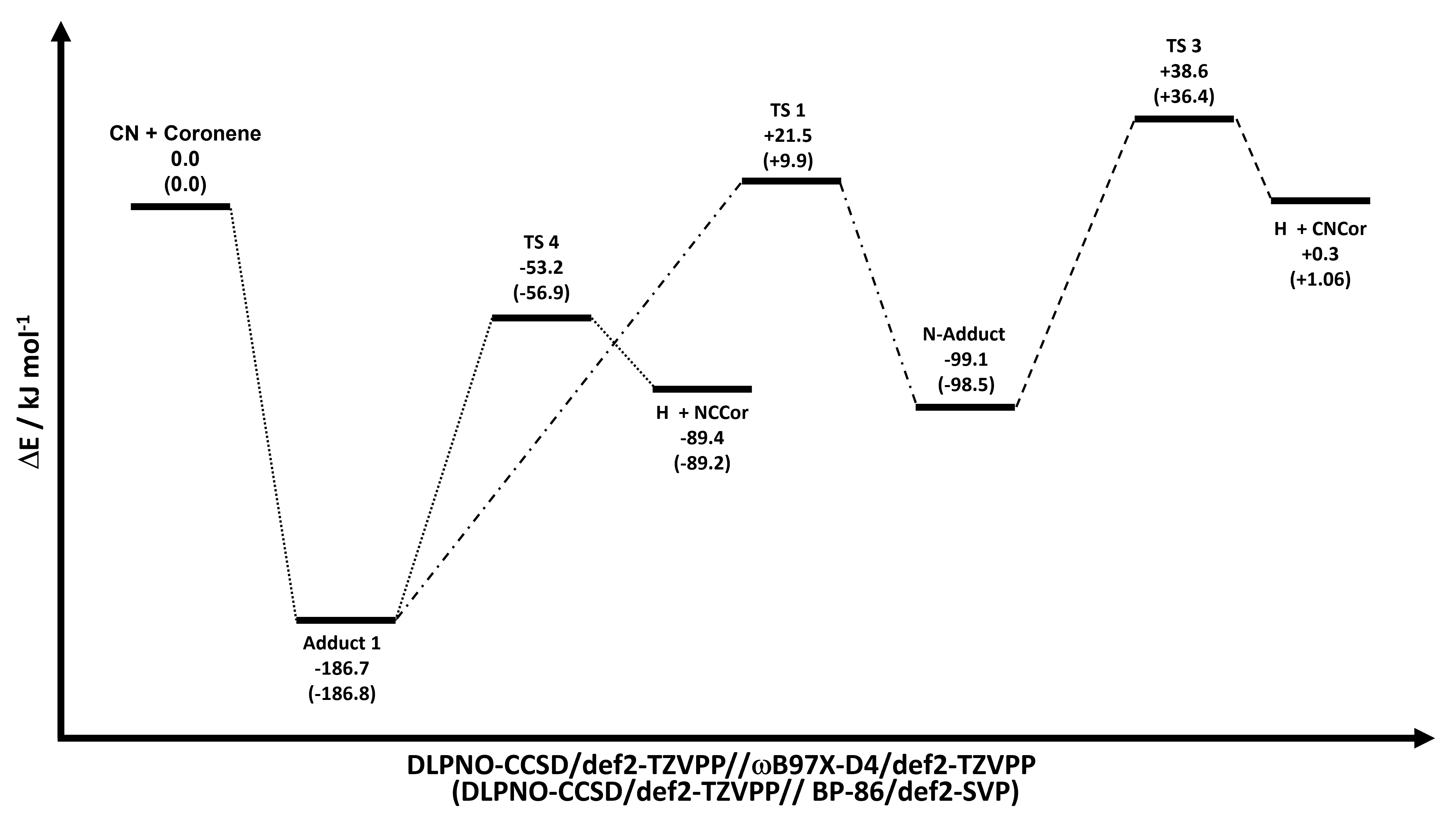}
    \caption{{Potential energy surfaces for the addition of \ce{CN} to coronene at the DLPNO-CCSD/def2-TZVPP//$\omega$B97X-D4/def2-TZVPP level with DLPNO-CCSD/def2-TZVPP//BP-86-D3(BJ)/def2-SVP level results presented in parenthesis. The upper frame provides the surface where a loose entrance channel complex is linked to the formation of the adducts. In the lower panel, this complex is neglected with barrierless approach forming the cyano adduct.}}
    \label{fig:PES}
\end{figure*}

A loose Van der Waals (VdW) complex was found using DFT-Cheap, which failed to re-optimize with DFT-2. Additional scans were carried out of the entrance channel with DFT-2 and although a shallow minima and submerged barrier were apparent on these, the shallow minima again did not optimize to loose complexes. Therefore, there is uncertainty in the presence of a loose entrance channel complex and as such two separate surfaces were treated in subsequent master equation calculations one where both adducts were linked to an entrance channel complex with energetics defined at the DLPNO-CCSD//DFT-Cheap level and the second where no entrance channel barriers were present for the formation of the cyanocoronene adducts with all energetics at the DLPNO-CCSD//DFT-2 level. These surfaces are presented in Fig.~\ref{fig:PES} and summarized in Table~\ref{tab:PES}.

The relative energies given in Table~\ref{tab:PES} for both the DFT-Cheap and DFT-2 surfaces are in very good agreement for the overall reaction energetics once DLPNO-CCSD corrections have been applied both with the aug-cc-pVDZ and def2-TZVPP basis sets. In addition, the relative energies of the cyano and isocyano adducts formed by addition of \ce{CN} to coronene and their subsequent barriers to elimination of an H atom are in excellent agreement (deviate by less than 1\,kcal). Comparing to canonical couple cluster results including single and double excitations and perturbative triples with the cc-pVDZ basis set the agreement for the overall energetics of the reaction are well reproduced by the DLPNO-CCSD results. Incorporation of basis set extrapolation through the EP3 scheme leads to a large deviation of around 15\,kJ\,mol$^{-1}$. Considering the whole set of results provided in Table~\ref{tab:PES} it is reasonable to infer that the formation of isocyanocoronene is slightly endothermic at 0\,K. The effect of spin contamination was considered by carrying out further DLPNO-CCSD/def2-TZVPP calculations with spin restricted open shell Hartree Fock (ROH-DLPNO-CCSD), as well as the spin unrestricted (DLPNO-CCSD) calculations with the def2-TZVPP basis set. These results typically deviated by less than 10\,kJ\,mol$^{-1}$. 

For both the UHF-DLPNO-CCSD//DFT-Cheap and UHF-DLPNO-CCSD//DFT-2 surfaces emerged barriers were found for both the formation of the isocyanocoronene adduct (+9.9 \& 21.5\,kJ\,mol$^{-1}$)  and the subsequent H atom elimination leading to the formation of isocyanocoronene (+36.4 \& +38.6\,kJ\,mol$^{-1}$). Whereas, the formation of the cyanocoronene occurs over a submerged barrier in the entrance channel (-18.6 \& -20.1\,kJ\,mol${-1}$) and the elimination of a H atom leading to the formation of cyanocoronene occurs over a very submerged barrier on the exit channel (-56.9 \& -53.2\,kJ\,mol$^{-1}$).

\begin{table}
   \caption{Relative energies for selected stationary points on the \ce{CN} + coronene potential energy surface including scaled harmonic zero point energies. In the column headers 1- represents using BP-86-D3(BJ)/def2-SVP structures and 2- using  $\omega$B97X-D4/def2-TZVPP structures. With the single point energy corrections applied being given as -1 for DLPNO-CCSD/def2-TZVPP, ROHF-DLPNO-CCSD/def2-TZVPP -3 for DLPNO-CCSD/aug-cc-pVDZ, -4 for CCSD(T)/cc-pVDZ and -5 for EP3-MP2 (DZ-QZ). All values are reported in kJ\,mol$^{-1}$.}
    \centering
    \begin{tabular}{lcccccccccc}
        \toprule
         &  1-1&   1-2&1-3& 1-4&  1-5&  2-1&   2-2&2-3& 2-4&  2-5\\
    \midrule
         \ce{CN} + Coronene&  0.0&   0.00
&0.0&  0.0&  0.0&  0.0&   0.00
&0.0&  0.0& 0.0
\\
         VdW&  -20.57&   -20.26
&-25.82&  &  &  *&   *
&*&  &  
\\
         TS 1 (CN addition)&  9.87&   14.42
&3.83&  &  &  21.48&   25.03
&13.18
&  & 
\\
         N-Adduct 1&  -98.48&   -90.90&-98.57&  &  &  -99.08&   -91.60
&-101.41
&  & 
\\
         TS 3 H elimination&  36.35&   42.57
&33.38&  &  &  38.55&   45.37
&34.23
&  & 
\\
         H + Isocyanocoronene&  1.06&   0.14
&1.72&  1.12&  -16.20&  0.33&   -0.68
&0.20
&  0.94& -14.17
\\
         TS 2 (NC addition)&  -18.57&   -18.27
&-23.82&  &  &  -20.09&   -18.52
&-25.05
&  & 
\\
         Adduct 1&  -186.80&   -179.23
&-184.50&  &  &  -186.68&   -179.22
&-186.00
&  & 
\\
         TS 4 H elimination &  -56.87&   -50.57
&-57.91&  &  &  -53.20&   -46.27
&-54.98
&  & 
\\
         H + Cyanocoronene&  -89.15&   -89.97
&-86.33&  -89.88&  -106.47&  -89.39&   -90.31
&-86.53
&  -88.90& -103.67
\\
\bottomrule
    \end{tabular}
 
    \label{tab:PES}
\end{table}

\subsection{Master equation kinetic predictions}
The UHF-DLPNO-CCSD//DFT-2 surface including and excluding the UHF-DLPNO-CCSD//DFT-Cheap loose entrance channel complex were incorporated into the energy-grained master equation calculator MESMER 7.1~\citep{Glowacki:2012:9545}, which allowed the reaction to be simulated over a range of densities ( $1\times 10^{4}$ to $1\times 10^{15}\,\mathrm{cm^{-3}}$) and temperatures (20 -- 3000\,K). Where the entrance channel complex was included in MESMER, simulations were primarily carried out with the UHF-DLPNO-CCSD corrected DFT-Cheap barriers to addition. However, repeats were carried out with the  UHF-DLPNO-CCSD//DFT-2 barrier to the formation of the isocyano-adduct (the UHF-DLPNO-CCSD//DFT-2 barrier to the formation of the cyano-adduct was submerged with respect to the UHF-DLPNO-CCSD//DFT-Cheap entrance channel complex). In addition, simulations were carried out where all the barriers were raised by 10\,kJ\,mol$^{-1}$ and where all the barriers were reduced by 10\,kJ\,mol$^{-1}$.  

The temperature-dependent collision rate coefficient for the barrierless formation of the cyano-adduct or the loose complex was estimated using classical capture theory (CCT)~\citep{Georgievskii:2005:194103}. 

\begin{equation}\label{eq:kcoll}
    k_{\textrm{coll}}(T) = \sigma_{\textrm{coll}}\langle v(T) \rangle = \left[\pi \left(\frac{2C_6}{k_BT}\right)^{1/3} \Gamma \left( \frac{2}{3}\right)\right] \left[ \left( \frac{8k_BT}{\pi \mu}\right)^{1/2} \right],\vspace{12pt}
\end{equation}
In equation~\ref{eq:kcoll}, $k_B$ is the Boltzmann constant, $\Gamma(x)$ is the gamma function such that $\Gamma(2/3) = 1.353$, $\mu$ is the reduced mass of the collision, and $C_6$ is the sum of coefficients describing the magnitude of the attractive forces between collision partners given as equation~\ref{eq:kcollpart2}.

\begin{equation}\label{eq:kcollpart2}
C_6=  \frac{2}{3}\left(\frac{\mu_1^2 \mu_2^2}{k_B T\left(4 \pi \epsilon_0\right)^2}\right)+\frac{\mu_1^2 \alpha_2+\mu_2^2 \alpha_1}{4 \pi \epsilon_0}  +\frac{3}{2} \alpha_1 \alpha_2\left(\frac{I_1 I_2}{I_1+I_2}\right) 
\end{equation}

In equation~\ref{eq:kcollpart2},  $\epsilon_0$ represents the permittivity of free space, $\mu_1$ and $\mu_2$ are the dipole moments of the reactants, and $\alpha_1$/$\alpha_2$ and $I_1$/$I_2$ are their polarizabilities and ionization energies, respectively. These parameters were taken from the online databases NIST Chemistry WebBook and CCCBDB~\citep{NISTwebbook,CCCBDB-2022}. This approach has been shown to be accurate within a factor of 2 for the prediction of rate coefficients for barrierless radical neutral reactions~\citep{West:2019:134}. The re-dissociation of the adducts was then treated with the inverse Laplace transformation (ILT) methodology~\citep{Davies:1986:373} in MESMER. The subsequent reactions through defined transition states were treated with Rice–Ramsperger–Kassel–Marcus (RRKM) theory ~\citep{Holbrook,Baer:1996:,Lourderaj:2009:2236} including quantum mechanical tunneling effects through a 1 dimensional asymmetric Eckart barrier~\citep{Miller:1979:6810}.

\begin{figure*}[bt]
    \centering
    \includegraphics[width=0.90\textwidth]{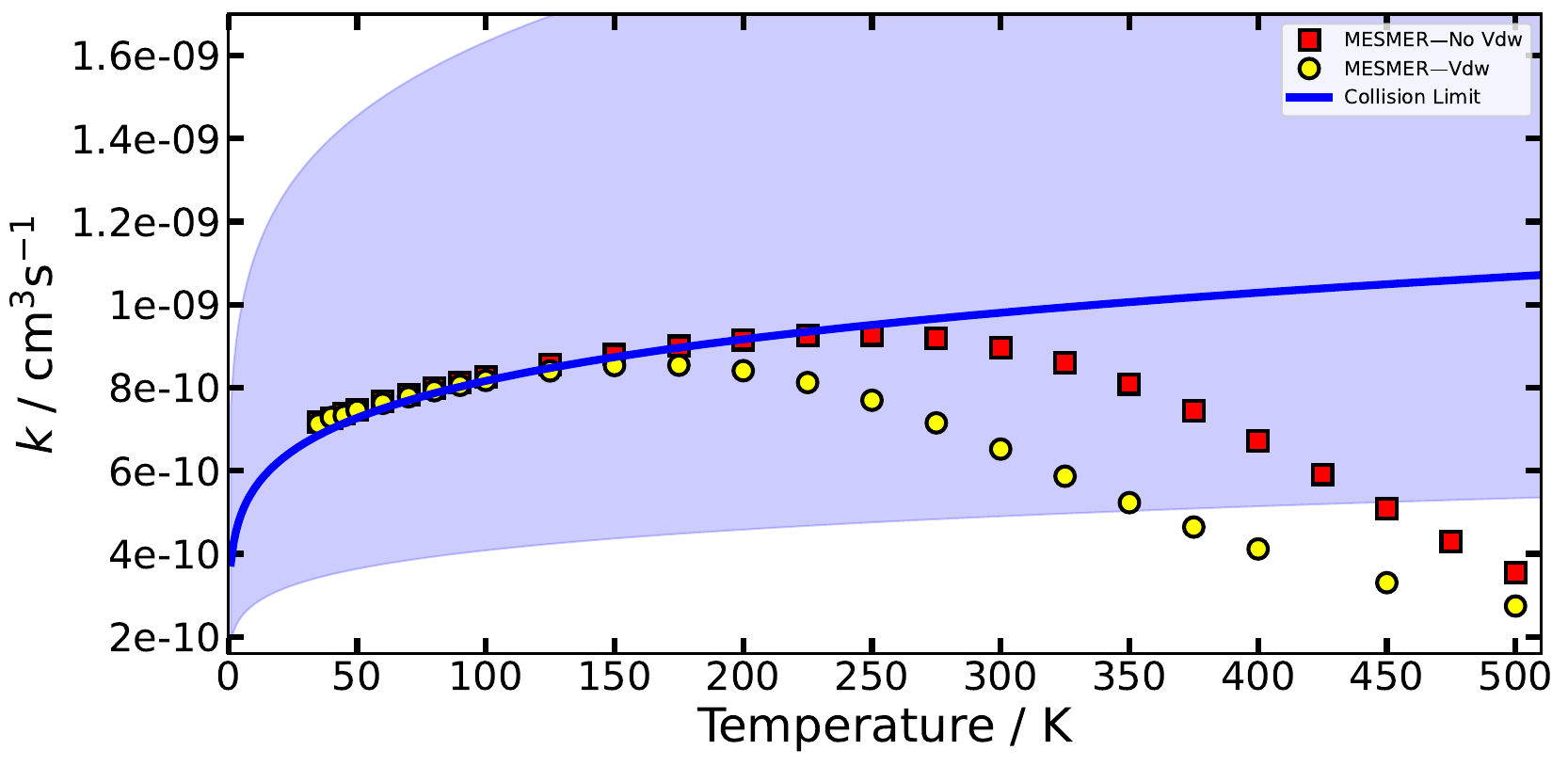}
    \caption{MESMER simulated bimolecular rate coefficients leading to the formation of H and cyanocoronene from the reaction of \ce{CN} with coronene with the CCT collision rate coefficient given as a solid blue curve; the estimated uncertainty in this is presented as the shaded region. The MESMER predictions including the loose complex on the entrance channel are given as yellow circles and those neglecting this are given as red squares.}
    \label{fig:MESMER}
\end{figure*}

Below 200\,K the results became sensitive to the grain size chosen with a grain size of 10\,cm$^{-1}$ by used for calculations under 200\,K even with a grain size of 2--5\,cm$^{-1}$ calculations failed below 20\,K. From 125\,K to 20\,K with a gas density of $2\times10^4\,\mathrm{cm^{-3}}$ in the complex included and from 200\,K in the complex excluded case the calculated bimolecular rate coefficients leading to the formation of cyanocoronene and an H atom from the reaction of \ce{CN} and coronene matched the estimated collision rate as can be seen in Fig.~\ref{fig:MESMER}. Therefore, it is reasonable to predict that at a gas density of $2\times10^4\,\mathrm{cm^{-3}}$ and at 10\,K this reaction will occur at the collision rate ($k_\mathrm{col} = 5.6^{+5.6}_{-2.8} \times 10^{-10}\,\mathrm{cm^{3} s^{-1}}$) and lead solely to the formation of cyanocoronene (isocyanocoronene is not predicted to be formed under these conditions). In these simulations, the temperature-dependent energy transfer parameter $<$$\Delta$E$>$$_d$ was set as 80 $\times$ (T/298)$^{1.0}$\,cm$^{-1}$ using the value assumed in~\citet{Georgievskii:2005:194103}; additionally, it should be noted the result that the collision limit had been reached below 100 K was not sensitive to this parameter when it was varied between 50 and 150 $\times$ (T/298)$^{1.0}$\,cm$^{-1}$. 

As the route to the formation of isocyanocoronene is endothermic and involves emerged barriers, both in the entrance channel and to the elimination of an H atom leading to product formation, this channel is predicted to be negligible. Varying the height of these barriers by 10\,kJ\,mol$^{-1}$ did not result in any observed isocyanocorone formation. Repeats were additionally performed where these barriers were shifted to match the deviation between the DLPNO-CCSD/def2-TZVPP corrected and EP3-MP2(DZ-QZ) corrected energetics ($\approx$ 15\,kJ\,mol$^{-1}$) the entrance channel barrier becomes submerged but the exit barrier remains emerged and the resulting prediction showed no formation of isocyanocoronene. The route to the formation of cyanocoronene involves barriers that are submerged both in the entrance channel and to the subsequent H atom elimination. The depth to which the barriers are submerged on the route to the formation of cyanocoronene means that this route remained rapid under the conditions relevant to TMC-1 even when the barriers were raised by 15\,kJ\,mol$^{-1}$.  

\end{document}